\def\editmode{0}
\def\reportmode{1}

\def\bibfilenames{WISENET}

\if\reportmode1
\documentclass[10pt,final,twocolumn]{IEEEtran}
\else
\documentclass[journal]{IEEEtran}
\usepackage{hyperref}
\fi

\if\editmode1  
\usepackage[backend=bibtex,style=alphabetic,sorting=debug,maxbibnames=99]{biblatex}
\DeclareFieldFormat{labelalpha}{\thefield{entrykey}}
\DeclareFieldFormat{extraalpha}{}
\bibliography{\bibfilenames}
\newcommand{\cmt}[1]{\noindent\textcolor{lightgreen}{\underline{[#1]}}} 
\newcommand{\hc}[1]{\textcolor{blue}{#1}} 
\newenvironment{myitemize}{\begin{itemize}}{\end{itemize}}
\newcommand{\myitem}{\item}

\else

\usepackage{bm}
\usepackage{cite}
\newcommand{\cmt}[1]{} 
\newcommand{\hc}[1]{\textcolor{black}{#1}} 
\newenvironment{myitemize}{}{}
\newcommand{\myitem}{}

\usepackage{flushend}
\fi 
\usepackage{amsmath,graphicx}
\usepackage{epstopdf}
\usepackage{multicol,multirow}
\usepackage{arydshln}

\usepackage[linesnumbered,ruled,vlined]{algorithm2e}

\usepackage{fixltx2e}

\usepackage{graphicx}

\usepackage{subcaption}              

\usepackage[utf8]{inputenc}

\usepackage{amsfonts}
\usepackage{amsmath}
\usepackage{mathtools}

\usepackage{amssymb}

\usepackage{bm}

\usepackage{color,verbatim}
\usepackage{multirow}
\usepackage{accents}

\usepackage{theoremref}

\usepackage{url}

\newcounter{rulecounter}
\newcommand{\resetrule}{ \setcounter{rulecounter}{0}}
\resetrule

\newsavebox{\selvestebox}
\newenvironment{colbox}[1]
  {\newcommand\colboxcolor{#1}%
   \begin{lrbox}{\selvestebox}%
   \begin{minipage}{\dimexpr\columnwidth-2\fboxsep\relax}}
  {\end{minipage}\end{lrbox}%
   \begin{center}
   \colorbox{\colboxcolor}{\usebox{\selvestebox}}
   \end{center}}

\definecolor{orange}{rgb}{1,0.8,0}
\definecolor{gray}{rgb}{.9,0.9,0.9}
\definecolor{darkgray}{rgb}{.3,0.3,0.3}
\definecolor{darkblue}{rgb}{.1,0.0,0.3}
\definecolor{lightblue}{rgb}{0.7,0.7,1}
\definecolor{lightred}{rgb}{1,0.7,.7}
\definecolor{purple}{RGB}{204,153,255}
\definecolor{lightgray}{rgb}{.95,0.95,0.95}
\definecolor{lightgreen}{rgb}{0.3,0.5,0.3}
\definecolor{darkgreen}{rgb}{0.05,0.3,0.05}

\newcommand{\ra}{$\rightarrow$~}

\newcommand{\brackets}[1]{\left\{#1\right\}}
\newcommand{\bbm}[1]{{\bar{\bm #1}}}

\newcommand{\hbm}[1]{{\hat{\bm #1}}}

\newcommand{\cfield}{\mathbb{C}}
\newcommand{\rfield}{\mathbb{R}}

\newcommand{\diag}[1]{\mathop{\rm diag}\brackets{#1}}

\newcommand{\rank}{\mathop{\rm rank}}

\newcommand{\sinc}{\mathop{\rm sinc}}

\newtheorem{myproposition}{Proposition}
\newtheorem{myremark}{Remark}
\newtheorem{myproblemstatement}{Problem Statement}
\newtheorem{mylemma}{Lemma}
\newtheorem{mytheorem}{Theorem}
\newtheorem{mydefinition}{Definition}
\newtheorem{mycorollary}{Corollary}

\newcommand{\arev}[1]{\textcolor{black}{#1}}
\begin{document}

\newcommand{\sqrtmo}{{\hc{\jmath}}}
\newcommand{\regpar}{{\hc{\lambda}}}
\newcommand{\evalregpar}{{\hc{\mu}}} 
\newcommand{\pilotsigmat}{\hc{\bm Y}}
\newcommand{\receivedpilotsig}{{\hc{y}}}

\newcommand{\sourcenum}{{\hc{L}}}
\newcommand{\sourceind}{{\hc{l}}}
\newcommand{\featurenum}{{\hc{M}}}
\newcommand{\featureind}{{\hc{m}}}
\newcommand{\vecSelec}{{\hc{\bm o }}}
\newcommand{\matrixSelec}{{\hc{\bm O}}}
\newcommand{\missingfeat}{\hc{\text{FiM}}}
\newcommand{\featurevec}{\hc{\bm \phi}}
\newcommand{\featuremat}{\hc{\bm \Phi}}
\newcommand{\featurerow}{\hc{\bm \varphi}}

\newcommand{\featurerowone}{\hc{\varphi}}
\newcommand{\featureset}{\hc{\mathcal{F}}}
\newcommand{\rkhs}{\hc{\mathcal{H}}}
\newcommand{\featurefunest}{{\hc{\hat d}}}
\newcommand{\featurefun}{{\hc{d}}}
\newcommand{\smoothlossfunc}{{\hc{f}}}
\newcommand{\completedfeaturemat}{\hc{\mathring{\bm \Phi}}}
\newcommand{\completedfeaturevec}{\hc{\mathring{\bm \phi}}}
\newcommand{\featurecoordinatemat}{\hc{\bbm \Phi}}

\newcommand{\newfeaturenum}{{\desiredrank}}
\newcommand{\newfeatureind}{{\hc{\bar m}}}
\newcommand{\featurecoordinatevec}{{\hc{\bbm \phi}}}
\newcommand{\missfeaturemat}{{\hc{ \breve{\bm \Phi}}}}
\newcommand{\testmissfeaturevec}{{\hc{ \breve{\bm \phi}}}}
\newcommand{\rotatedfeaturevec}{{\hc{\bm \phi'}}}
\newcommand{\rotatedfeaturemat}{{\hc{\bm \Phi'}}}
\newcommand{\newfeaturemat}{{\hc{\bbm \Phi}}}

\newcommand{\pilotsig}{{\hc{a}}}
\newcommand{\noisesig}{{\hc{w}}}

\newcommand{\threshold}{{\hc{\gamma}}}
\newcommand{\sensitivity}{{\hc{\Gamma}}}
\newcommand{\retainedvar}{{\hc{\eta}}}
\newcommand{\desiredrank}{{\hc{r}}}
\newcommand{\obsfeaturenum}{{ \hc{ \breve{M}}}}
\newcommand{\obsfeatureind}{{ \hc{ \breve{m}}}}

\newcommand{\toa}{{\hc{\tau}}}
\newcommand{\toavec}{{\hc{\bm\tau}}}
\newcommand{\tdoa}{{\hc{\Delta}}}

\newcommand{\com}{\hc{\text{CoM}}}

\newcommand{\lagind}{{\hc{i}}}

\newcommand{\sensorsourcesourcenot}[3]{_{#2,#3,#1}}
\newcommand{\egmpar}{{\hc{\zeta}}}
\newcommand{\covFeatures}{\hc{\bm C}}
\newcommand{\Cov}{\mathrm{Cov}}
\newcommand{\trace}{\mathrm{Tr}}
\newcommand{\bigzero}{\mbox{\normalfont\Large 0}}

\title{Location-free Spectrum Cartography}

\if\reportmode1
 \author{Yves Teganya,~\IEEEmembership{Student Member,~IEEE,} Daniel Romero,~\IEEEmembership{Member,~IEEE,}\\ Luis Miguel Lopez Ramos,~\IEEEmembership{Member,~IEEE,}  
  and Baltasar Beferull-Lozano,~\IEEEmembership{Senior Member,~IEEE}
\thanks{The work in this paper was supported by the FRIPRO TOPPFORSK grant 250910/F20 and the INFRASTRUCTURE ReRaNP grant 245699/F50 from the Research Council of Norway.
 }
\thanks{The authors  are with the WISENET Lab, Dept. of ICT,
   University of Agder, Jon Lilletunsvei 3, Grimstad, 4879 Norway. E-mails:\{yves.teganya, daniel.romero, luismiguel.lopez, baltasar.beferul\}@uia.no.}
\thanks{Parts of this work have been presented in ICCASP 2018~\cite{teganya2018localization}.}
}
\else
\author{Yves Teganya,~\IEEEmembership{Student Member,~IEEE,} Daniel Romero,~\IEEEmembership{Member,~IEEE,}\\ Luis Miguel Lopez Ramos,~\IEEEmembership{Member,~IEEE,}  
  and Baltasar Beferull-Lozano,~\IEEEmembership{Senior Member,~IEEE}
\thanks{The work in this paper was supported by the FRIPRO TOPPFORSK grant 250910/F20 and the INFRASTRUCTURE ReRaNP grant 245699/F50 from the Research Council of Norway.
 }
\thanks{The authors  are with the WISENET Lab, Dept. of ICT,
   University of Agder, Jon Lilletunsvei 3, Grimstad, 4879 Norway. E-mails:\{yves.teganya, daniel.romero, luismiguel.lopez, baltasar.beferul\}@uia.no.}
\thanks{Parts of this work have been presented in ICCASP 2018~\cite{teganya2018localization}.}
}
\fi

\maketitle
\begin{abstract}
Spectrum cartography constructs maps of metrics such as channel gain
or received signal power across a geographic area of interest using
spatially distributed sensor measurements. Applications of these
maps include network planning, interference coordination, power
control, localization, and cognitive radios to name a few. Since
existing spectrum cartography techniques require accurate estimates of
the sensor locations, their performance is drastically impaired by multipath  affecting the
positioning pilot signals, as occurs in
indoor or dense urban scenarios. To overcome such a limitation, this
paper introduces a novel paradigm for spectrum cartography, where
estimation of spectral maps relies on features of these positioning
signals rather than on location estimates. Specific learning
algorithms are built upon this approach and offer a markedly improved
estimation performance than existing approaches relying on
localization, as demonstrated by  simulation studies in  indoor
scenarios.
\end{abstract}

\if\reportmode0
\fi

\section{Introduction}
\label{sec:introduction}

\cmt{Motivation}
\begin{myitemize}
\myitem\cmt{cartography overview}Spectrum cartography constructs maps of a certain channel metric, such as received signal
power, power spectral density (PSD), or channel gain over a geographical area
of interest by relying on measurements collected by radio frequency (RF) sensors~\cite{alayafeki2008cartography,bazerque2010sparsity,jayawickrama2013compressive}.
\myitem\cmt{motivating applications}%
\begin{myitemize}\myitem\cmt{communications}The obtained maps are of
utmost interest in a number of tasks in  wireless communication networks,  such as
network planning, interference coordination, power control, and
dynamic spectrum access~\cite{romero2017spectrummaps,grimoud2010rem,
dallanese2011powercontrol}.
\begin{myitemize}%
\myitem\cmt{net.planning}For instance, power maps can be useful in
network planning since the former indicate areas of weak coverage,
thus suggesting locations where new base stations must be deployed. 
\myitem\cmt{interf.coord}Since PSD maps characterize the
distribution of the RF signal power per channel over
space, they can play a major role in increasing frequency reuse to
mitigate interference.
\myitem\cmt{handoff}These maps  may also be of 
interest to speed up hand-off in cellular networks since they enable 
mobile users to determine the power of all channels at a given
location without having to spend time measuring it.
\myitem\cmt{DSA}Additional use cases may include
cognitive radios, where secondary users aim at exploiting  underutilized spectrum
resources in the space-frequency-time domain, or
\end{myitemize}%
\myitem\cmt{source localization}source
localization, where the locations of certain
transmitters may be estimated by inspecting a map~\cite{bazerque2010sparsity}.     
\end{myitemize}
\end{myitemize}

\begin{figure}
\centering
\begin{subfigure}{4cm}
\centering\includegraphics[width=4cm]{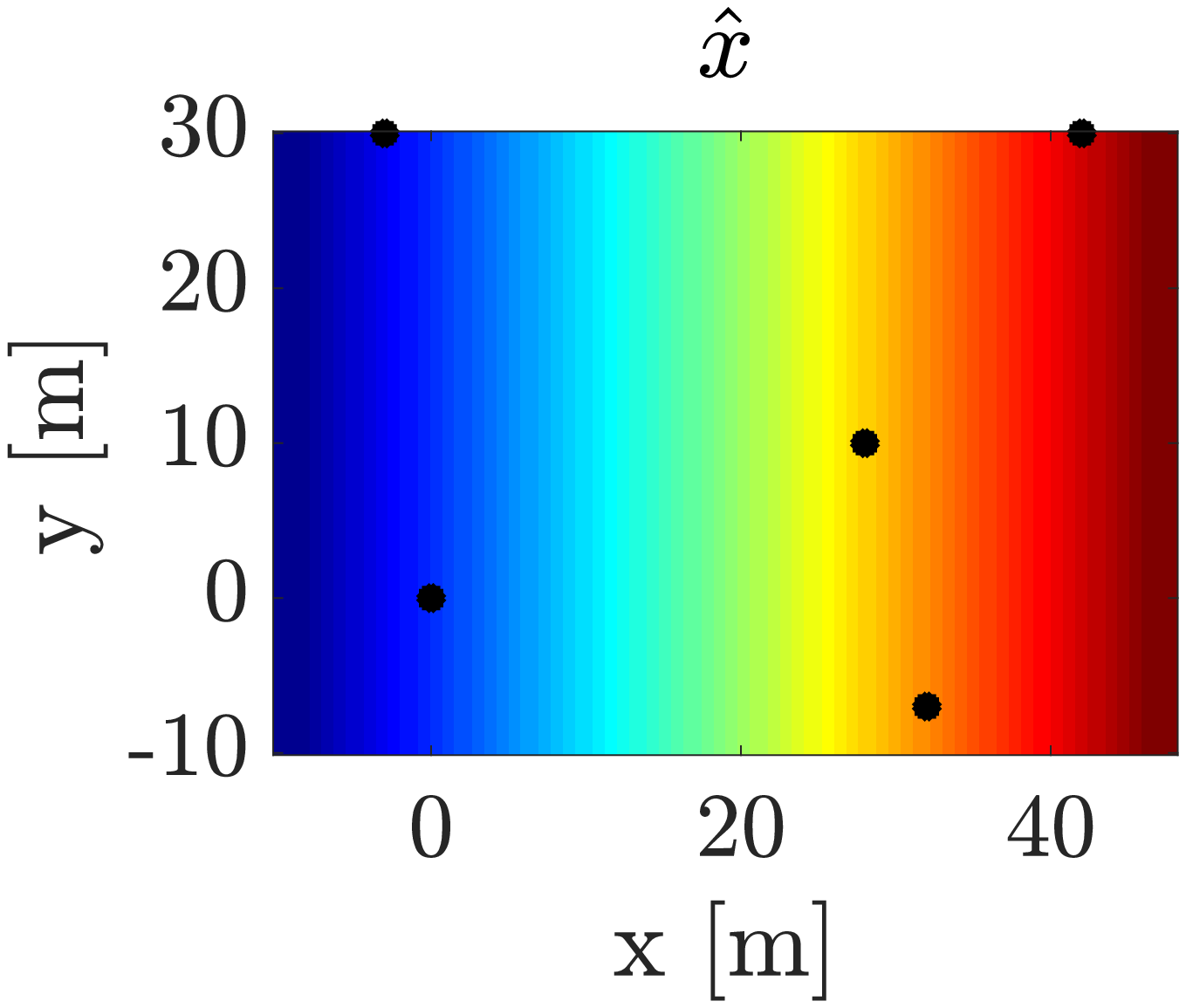}
\caption{}
\end{subfigure}%
\begin{subfigure}{4cm}
\centering\includegraphics[width=4cm]{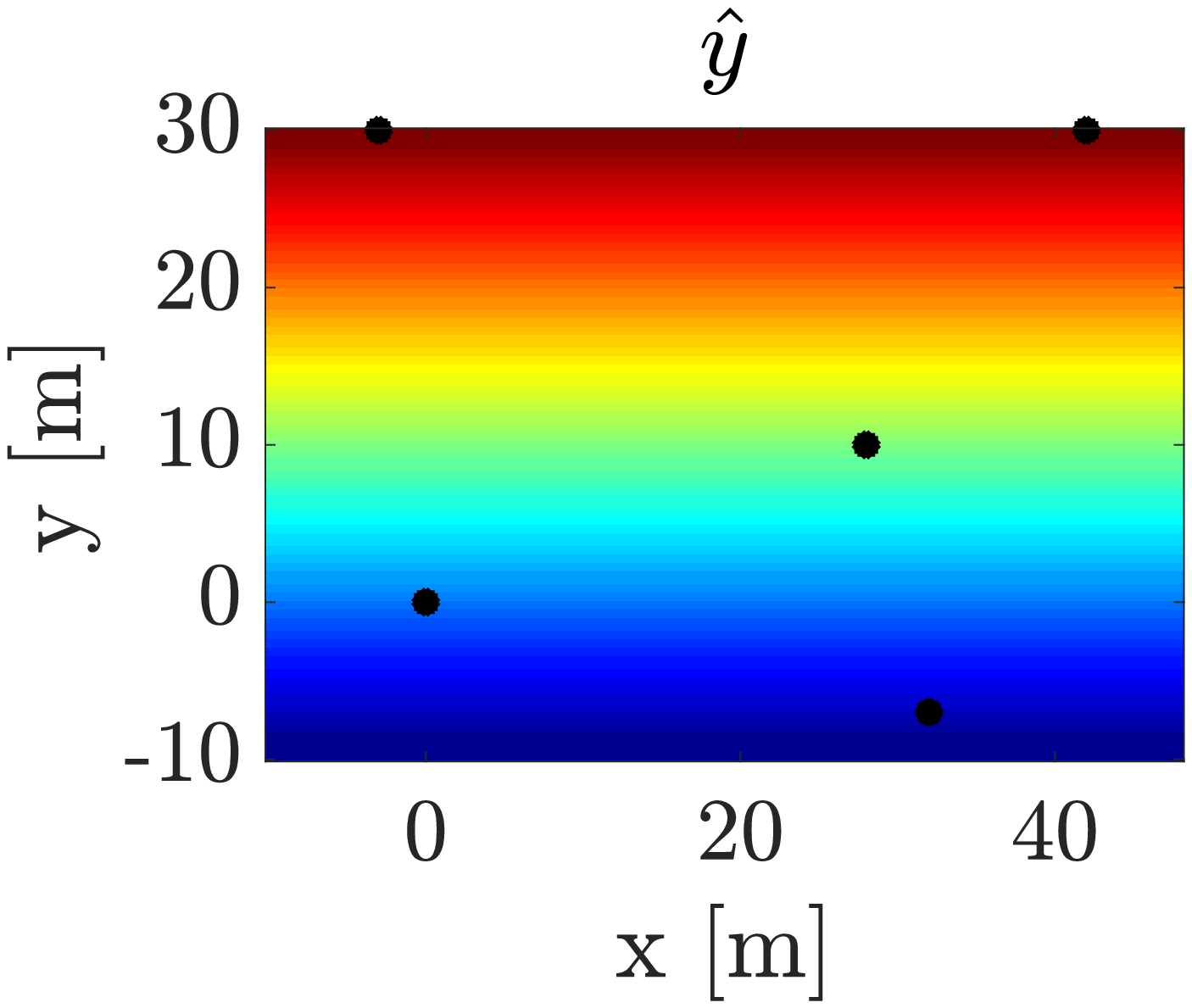}
\caption{}
\end{subfigure}\vspace{0pt}
\begin{subfigure}{4cm}
\centering\includegraphics[width=4cm]{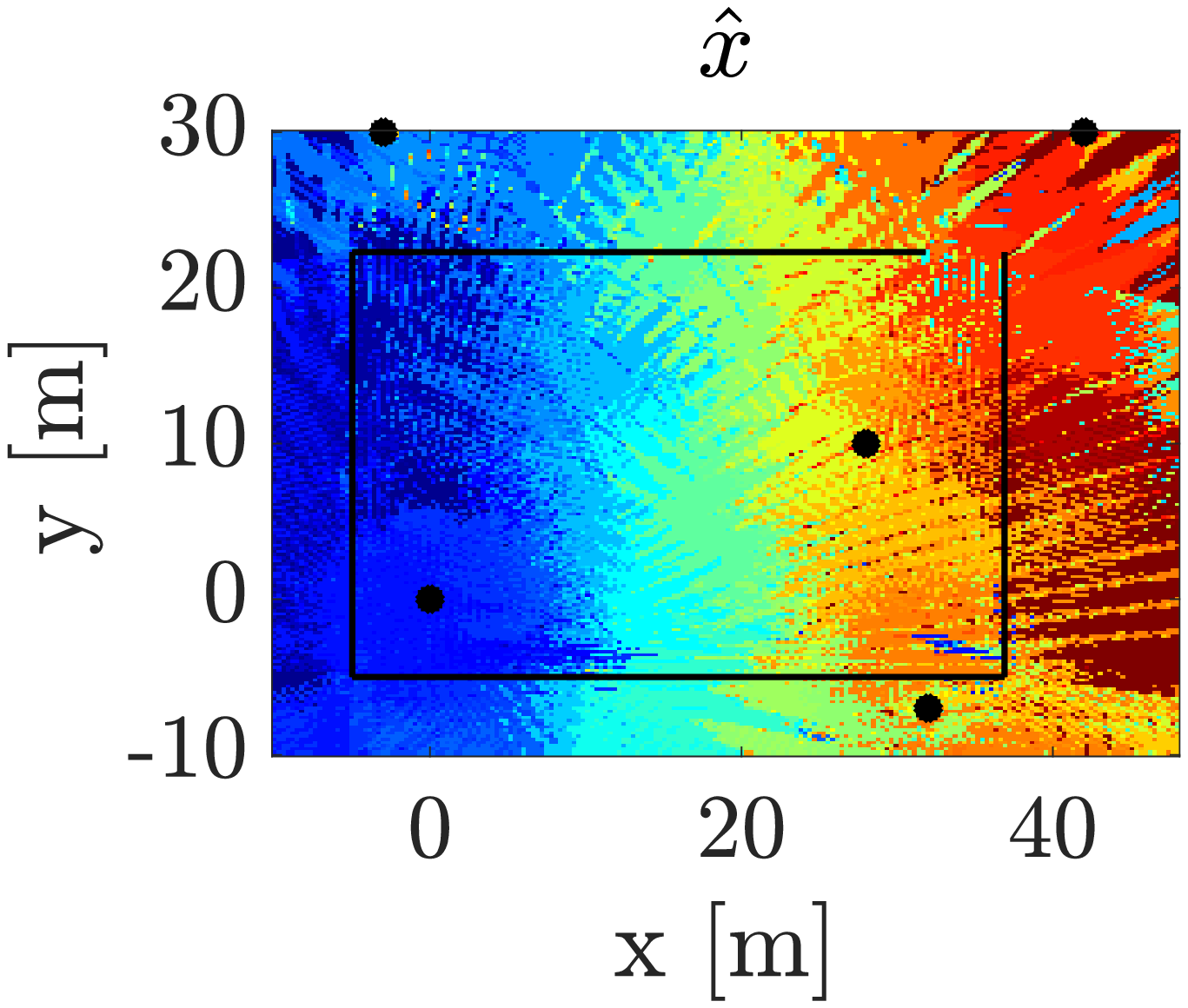}
\caption{}
\end{subfigure}%
\begin{subfigure}{4cm}
\centering\includegraphics[width=4cm]{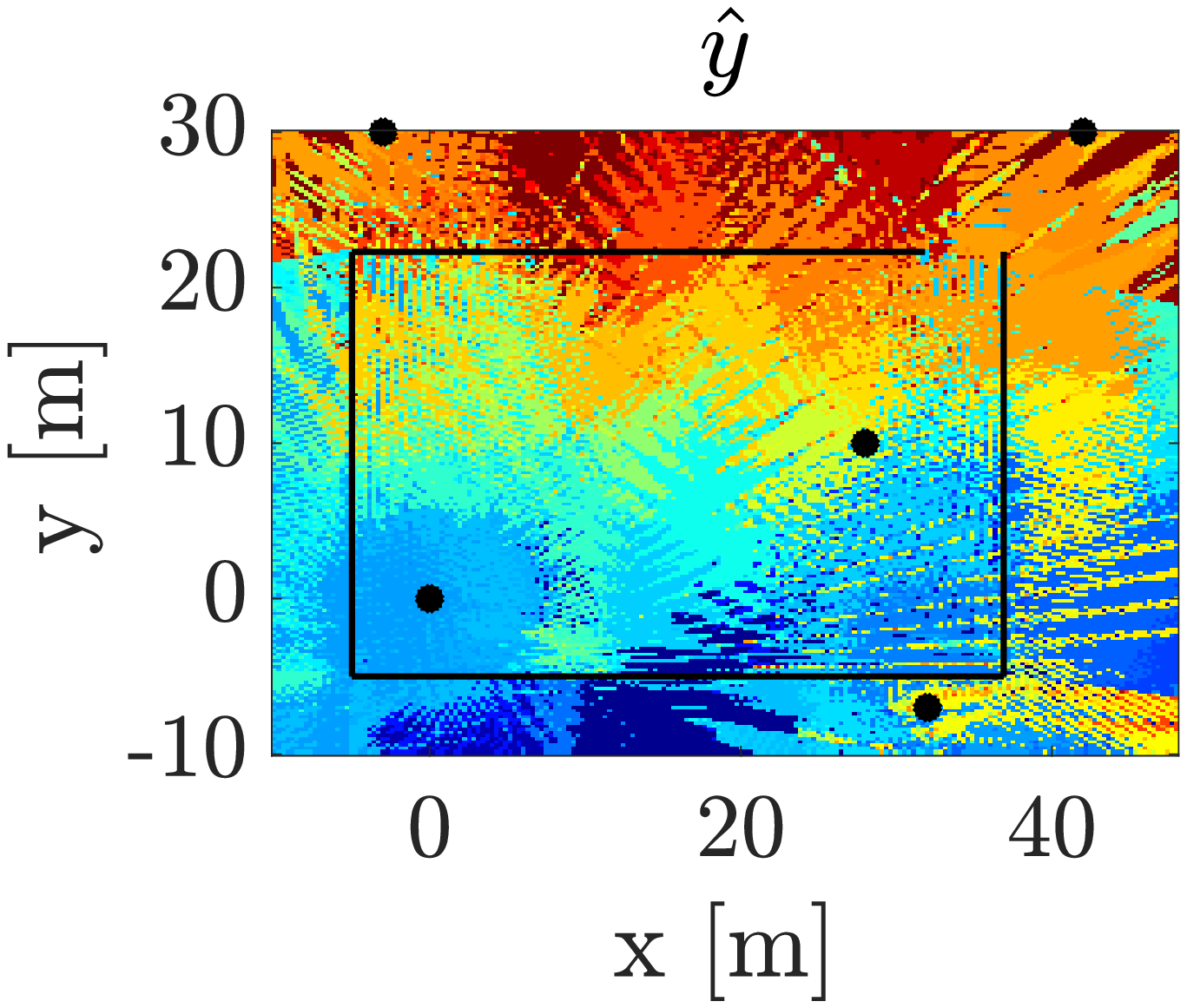}
\caption{}
\end{subfigure}
\caption{ Estimation of spatial coordinates using TDoA: (a)-(b) in free space, (c)-(d) indoor where the solid black lines represent the walls of the building; the black dots represent the locations of the anchor base stations. \arev{The color of each pixel represents the value of the estimated location coordinate at each point in the $150\times 150$ grid area. Because location estimates in (a)-(b) coincide with the true locations, they can act as colorbars to the estimates in (c)-(d).}} 
\label{f:locEst}
\end{figure}

\cmt{Literature review}
\begin{myitemize}
\myitem\cmt{power maps}Existing methods for mapping  RF power apply spatial 
interpolation or regression techniques to power measurements collected by
spatially distributed sensors.
\begin{myitemize}%
\myitem\cmt{kriging}Some of these methods include kriging~\cite{van2004kriging,alayafeki2008cartography,boccolini2012wireless},
\myitem\cmt{compress. sensing, dict. lear., matrix
completion}orthogonal matchning pursuit~\cite{jayawickrama2013compressive}, matrix
completion~\cite{ding2016cellular},
dictionary learning~\cite{kim2011link,kim2013dictionary},
\myitem\cmt{bayes. models}sparse Bayesian
learning~\cite{huang2015cooperative},
\myitem\cmt{kernel}or kernel-based learning~\cite{bazerque2013basispursuit,hamid2017non}.
\end{myitemize}%
\myitem\cmt{PSD maps}%
\begin{myitemize}%
\myitem\cmt{sparsity}Since these
works can only map power distribution across space but not across
frequency, different schemes have been devised to construct PSD maps,
for instance by exploiting the sparsity of power
distributions over space and frequency with a basis expansion
model~\cite{bazerque2010sparsity,bazerque2011splines}
\myitem\cmt{kernel methods}or by leveraging the
framework of
kernel-based learning~\cite{romero2017spectrummaps}.
\end{myitemize}%
\myitem\cmt{other metrics \ra channel gain maps}Rather than mapping
power, other families of methods construct channel-gain maps
using
\begin{myitemize}%
\myitem\cmt{KKF}Kriged Kalman filtering~\cite{kim2011cooperative},
\myitem\cmt{RKHS}non-parametric regression in reproducing kernel Hilbert spaces (RKHSs)~\cite{romero2016blindchannelgain},
\myitem\cmt{Low rank and sparsity}low rank and sparsity~\cite{lee2016lowrank}, 
\myitem\cmt{Markov RF}or hidden Markov random fields~\cite{lee2018adaptive}. 
\end{myitemize}
\myitem\cmt{Limitations}
\begin{myitemize}

\myitem\cmt{Need Locations}All the aforementioned
schemes require accurate knowledge of the sensor locations. For
this reason, they will be collectively referred to
as \emph{location-based (LocB) cartography}.
\myitem\cmt{Location not konwn}However, location is seldom known in
practice and therefore must be estimated from features such as the  received signal strength,
the time (difference) of arrival (T(D)oA), or the direction of 
arrival (DoA)  of positioning pilot signals transmitted by
satellites (e.g. in GPS)  or
terrestrial base stations (e.g. in LTE or 
WiFi~\cite{bshara2010fingerprinting})~\cite{naidu2017distributed, bensky2016wireless}.
\myitem\cmt{Location not
accurate in multipath env.}Unfortunately, accurate location estimates
are often not available in practice due to propagation phenomena
affecting those pilot signals such as multipath, which limits the
applicability of existing cartography techniques, especially in indoor
and dense urban scenarios.
\begin{myitemize}%
\myitem\cmt{Fig.}To see the intuition behind this observation, Figs.~\ref{f:locEst}a and~\ref{f:locEst}b
respectively show the
$x$ and $y$ coordinates of the location estimates obtained by applying a state-of-the-art
 localization algorithm to TDoA measurements of 5 pilot signals
 received in free space (details of the specific simulation setting can be
 found in Sec.~\ref{sec:numericalTest}). On the other hand,
 Figs.~\ref{f:locEst}c
 and~\ref{f:locEst}d depict the same estimates but in an indoor propagation
 scenario. As observed,  the estimates in the second case
are neither accurate nor smooth across space, which precludes any
reasonable estimate of a spectrum map based on them.

\myitem\cmt{other reasons of inaccuracy of location \ra expensive}\arev{To counteract this difficulty, there are three main types of indoor positioning systems~\cite{liu2007survey}}: 
\begin{myitemize}%
\myitem\cmt{uwb}\arev{(i) Those based on ultra-wideband (UWB)~\cite{uwb, yang2004ultra, dardari2008threshold}, which require a dedicated infrastructure and relatively high costs,  e.g. synchronized anchor nodes in the area where the map has to be constructed. Therefore, localization cannot be carried out in an area where such hardware is not present.}
\myitem\cmt{fingerprinting}\arev{(ii)
Other indoor positioning systems are based on fingerprinting~\cite{brunato2005statistical, prasithsangaree2002indoor, liu2007survey}, which involves a manual collection and storage of a dataset. This dataset may comprise the measured power of multiple  beacons at a set of known locations. Note that this process is time consuming and typically expensive because a human or robot should physically go through several known locations to take measurements. Furthermore, if there are significant changes in the propagation environment, these methods would require the acquisition of a new dataset.}
 \myitem\cmt{other IPS}\arev{(iii) There exist other indoor positioning systems that combine UWB or fingerprinting with ultrasound~\cite{HPsmartlocus} or RFID~\cite{ni2003landmarc}. Thus, they inherit the limitations of (i) and (ii) and require furthermore special sensors and/or line-of-sight propagation conditions.}
\end{myitemize}%
\end{myitemize}%
\myitem\cmt{bottom line}\arev{To sum up, all existing cartography schemes require accurate location information, which is not available in dense multipath and indoor scenarios when there are no special localization infrastructure or fingerprinting datasets.}  
\end{myitemize}
\end{myitemize}

\cmt{Contribution}
\begin{myitemize}
\myitem\cmt{idea: bypass localization \ra paradigm shift}The main contribution of this
paper is to address this limitation by proposing the framework of \emph{location-free (LocF)
cartography}. The key observation is that inaccurate location estimates introduce
significant errors in  spectrum map estimation. 
\myitem\cmt{how}To bypass this limitation, the
proposed approach obtains spectrum maps indexed directly by (or as a
function of) features of the received pilot signals.
\myitem\cmt{learning}\arev{Although many algorithms can be devised within this framework,}
\begin{myitemize}%
\myitem\cmt{kbl\ra simple}\arev{the present paper develops an algorithm based on kernel-based learning for the sake of exposition}. This is not only because of the
  simplicity, 
  flexibility, and good performance of kernel-based estimators, but also
  because they have well-documented merits in spectrum
  cartography~\cite{bazerque2011splines,romero2017spectrummaps}.
\myitem\cmt{frequency}\arev{Similarly, the discussion focuses on
  constructing power maps, but the proposed paradigm carries over to
  other metrics such as PSD.} 
\end{myitemize}%
\myitem\cmt{lower complexity and cheap}Remarkably, as a byproduct of skipping the
localization step,
\begin{myitemize}%
\myitem\cmt{lower compl}the resulting cartography algorithm is typically computationally less expensive than its LocB counterparts
\myitem\cmt{cheap}\arev{and does not require additional localization infrastructure or the costly creation of fingerprinting datasets.}
\end{myitemize}%
\myitem\cmt{Extensions}%
\begin{myitemize}%
\myitem\cmt{Feature design}The second main contribution is a  design of
pilot signal features tailored to multipath environments.
\myitem\cmt{Accommodate missing features}The third contribution is a
special technique to accommodate scenarios where a sensor can only
extract a subset of those  features due to low signal-to-noise ratio (SNR).
\end{myitemize}%
\myitem\cmt{Empirical validation}Finally, the proposed LocF
  cartography scheme is studied through Monte Carlo simulations in
  realistic propagation environments.
\myitem\cmt{caveat}As expected, the proposed scheme
  outperforms LocB cartography in multipath scenarios, but traditional LocB
  approaches are still preferable when  accurate
  location estimates are available. 

\end{myitemize}

\cmt{Paper organization}
The rest of this paper is structured as follows:
Sec.~\ref{sec:problemformulation} describes the system model, states
the problem, and reviews LocB
cartography. Sec.~\ref{sec:locfree} introduces  LocF cartography along
with the proposed map estimation algorithm, whereas
Sec.~\ref{sec:features} deals with feature design. Numerical tests are
presented in Sec.~\ref{sec:numericalTest}, and conclusions 
in Sec.~\ref{sec:conclusions}.

\cmt{Notation} \textit{Notation}: Scalars are denoted by lowercase
letters. Bold uppercase (lowercase) letters denote matrices (column
vectors), $\bm{I}_N$ is the $N \times N$ identity matrix and $\bm
1$ is the vector of all ones of appropriate dimension. The symbol
$\sqrtmo:=\sqrt{ -1}$ is the imaginary unit, ${(\cdot)^{*}}$ stands for the complex conjugate, while $\ast$ denotes  convolution. Furthermore, operators $(\cdot)^{\top}$ and $\vert \vert \cdot \vert \vert_F$ represent transposition and the Frobenius norm, respectively.

\section{Problem Formulation and LocB Cartography}
\label{sec:problemformulation}

\cmt{overview}This section formulates the general spectrum cartography
problem and reviews the
basics of LocB~cartography.

\begin{myitemize}
\myitem\cmt{system model}F
\begin{myitemize}
\myitem\cmt{Region and channel to map}The goal is to determine the
power $p(\bm {x})$ of a certain channel,
termed \emph{channel-to-map} (C2M), at every location $\bm
{x}\in\mathcal{X}$ of a geographical region of interest
$\mathcal{X}\subset\rfield^d$, with $d=2$ or $3$. \arev{For example, this C2M can be an uplink or downlink channel of a cellular network as well as a radio or TV broadcasting channel}.
\myitem\cmt{s}To this end, a collection of sensors gather $N$ measurements at
locations $\{\bm x_n\}_{n=1}^N\subset\mathcal{X}$  not necessarily known.
\myitem\cmt{measurements}The noisy measurement of the power $p(\bm x_n)$ at 
location $\bm x_n$ will be represented as 
$\tilde{p}_n$.
\myitem\cmt{no. sensors}Since the sensors collect measurements at multiple locations in 
$\mathcal{X}$,  the number of measurements may
be significantly greater than the number of sensors.
\end{myitemize}

\myitem\cmt{LocB cartography}
\begin{myitemize}
\myitem\cmt{location known}
\begin{myitemize}
\myitem\cmt{map estimation}In LocB cartography~\cite{alayafeki2008cartography,boccolini2012wireless,jayawickrama2013compressive,
kim2011link,kim2013dictionary,ding2016cellular,huang2015cooperative,hamid2017non,
bazerque2010sparsity,bazerque2011splines,romero2017spectrummaps,romero2016blindchannelgain,
lee2016lowrank,lee2018adaptive}, a fusion
center is ideally given  pairs $\{(\bm x_n,
\tilde{p}_n)\}_{n=1}^N$, which include the exact sensor locations $\{\bm x_n\}_{n=1}^N$,
and obtains a function estimate $\hat p(\bm {x})$ that provides the
power of the C2M at any query location $\bm {x}\in\mathcal{X}$.
\myitem\cmt{evaluation}%
With this function, a node at location $\bm {x}$ can determine the power of
the C2M if it knows $\bm {x}$.
\end{myitemize}%
\myitem\cmt{location unknown}%
\begin{myitemize}%
\myitem\cmt{pilot signals}In practice, however, location is typically unknown and hence
the  sensor at the $n$-th measurement point must estimate $\bm x_n$ by relying on pilot signals
$\{\receivedpilotsig_{\sourceind,n}[k]\}_{\sourceind=1}^\sourcenum$, where $\receivedpilotsig_{\sourceind, n}[k]$ denotes the $k$-th
sample of the pilot signal transmitted by the  $\sourceind$-th base
station\footnote{Although the discussion assumes for simplicity that the pilot
signals are transmitted by terrestrial base stations,  the proposed
scheme can also be applied when these pilot signals are transmitted by
satellites.} and
received at the $n$-th measurement point. For convenience, form the $\sourcenum\times K$
matrix $\pilotsigmat_n$ whose $(\sourceind,k)$-th entry is
$\receivedpilotsig_{\sourceind,n}[k]$. Note that these pilot signals are
generally transmitted through a separate channel, not necessarily
the~C2M. However, both channels may coincide, as it occurs in certain
cellular communication standards.  

\myitem\cmt{overall procedure}%
\begin{myitemize}%
\myitem\cmt{measurements}From  $\pilotsigmat_n$, the sensor at the
$n$-th  measurement point obtains the  estimate
$\hbm x_n :=\hbm x (\pilotsigmat_n)$ of $\bm x_n$  by means of some 
localization algorithm~\cite{naidu2017distributed, bensky2016wireless}.
\myitem\cmt{estimation}A fusion center then uses $\{(\hbm
x_n,\tilde{p}_n)\}_{n=1}^N$ to obtain an estimate $\hat p(\bm x)$ of the
function $p(\bm x)$. Therefore, if the location estimates $\{\hbm
x_n\}_{n=1}^N$ are noisy, so will be $\hat p(\bm x)$.
\myitem\cmt{evaluation}If a node at an unknown query location
wishes to determine the power of the C2M, it will use the pilot signals
$\pilotsigmat$ to obtain an estimate $\hbm {x}:=\hbm x(\pilotsigmat)$ of
its location and will evaluate the map estimate as $\hat p(\hbm
{x})$. In this case, $\pilotsigmat$ is a matrix whose $(\sourceind,k)$-th entry is given
by the $k$-th sample of the $\sourceind$-th pilot signal $\receivedpilotsig_{\sourceind}[k]$ at the
query location. Thus, such an estimation has two sources
of error: first, the location estimation error in $\hbm {x}$ and,
second, the map estimation error in $\hat {p}( \bm{x})$.
\end{myitemize}
\end{myitemize}
\end{myitemize} 

\end{myitemize}

\begin{myremark}One may argue that a node
can determine the power of the C2M at its location more efficiently by
measuring it rather than by receiving the pilot signals, applying a
localization algorithm, and evaluating the map.  Whereas this may be the
case for a single C2M, if the aim is to determine the PSD, the power
of many C2Ms, or the impulse
response, then the associated measurement time may be prohibitive,
which favors the adoption of spectrum cartography approaches.
\end{myremark}

\section{Location-Free Cartography }
\label{sec:locfree}
\cmt{overview}This section proposes LocF cartography,
which bypasses the localization step involved in all existing
cartography approaches. To this end, the LocF cartography
problem is formulated as a function estimation task in
Sec.~\ref{sec:composition} and solved via kernel-based learning in
Sec.~\ref{sec:kernel}.

\subsection{Map Estimate as a Function Composition}
\label{sec:composition}

\cmt{overview}As detailed in the previous section, existing spectrum
cartography techniques are heavily impaired by localization errors since
the maps they construct are functions of  noisy location
estimates. The main idea of the
proposed framework is to bypass such a dependence.  

\cmt{derivation LocF}

\newcommand{\pilotfunest}{\hc{\hat p_{\pilotsigmat}}}

\begin{myitemize}
\myitem \cmt{LocB as composition}To this end, it is worth
interpreting LocB cartography from a  more abstract perspective.
\begin{myitemize}
\myitem\cmt{composition}As detailed in
Sec.~\ref{sec:problemformulation}, the LocB map estimate is of the
 form $\hat p(\hbm x)$ with $\hbm x := \hbm x(\pilotsigmat)$ denoting
the output of the selected localization algorithm when the pilot
signals are given by
 $\pilotsigmat\in\mathcal{Y}$. Thus, this estimate can be seen as  a
 function of $\pilotsigmat$, i.e.  $\pilotfunest(\pilotsigmat):=\hat
 p(\hbm x(\pilotsigmat))$, which can be expressed schematically as:
\begin{alignat}{3}
\begin{aligned}  \label{eq:locasedComp}
&\mathcal{Y} & \overset{\hat{\bm x}}{\xrightarrow{\hspace*{0.7cm}}}
&\quad \mathcal{X} & \overset{\hat p}{\xrightarrow{\hspace*{0.7cm}}}& \quad  \mathbb{R}
\\
&\pilotsigmat &  \xrightarrow{\hspace*{0.7cm}}  & \quad \hat{\bm
x}(\pilotsigmat)& \xrightarrow{\hspace*{0.7cm}}& \quad  \hat p(\hbm x(\pilotsigmat)).
\end{aligned}
\end{alignat}
\myitem\cmt{estimation}As mentioned in
Sec.~\ref{sec:problemformulation}, existing (LocB)
cartography approaches obtain an estimate $\hat p$ of $p$ using the
data $\{(\hbm x(\pilotsigmat_n),\tilde{p}_n)\}_{n=1}^N$ for instance by searching
for a function in an RKHS~\cite{bazerque2013basispursuit,romero2017spectrummaps,
hamid2017non}. 
\myitem\cmt{regularly good}When $\hbm x(\pilotsigmat)$ is a reasonable
estimate of the location $\bm x$ at which $\pilotsigmat$ has been observed, such a LocB approach works well.
\myitem\cmt{limitations \ra too simplistic}However, due to multipath
propagation effects impacting the pilot signals in $\pilotsigmat$, $\hbm
x(\pilotsigmat)$ may be very different from $\bm x$, which drastically hinders the estimation of $p$. Thus, in those
cases where the location estimates $\{\hbm x(\pilotsigmat_n)\}_{n=1}^N$ are
noisy, the resulting estimate $\hat p$, and consequently $\pilotfunest$,
will be correspondingly noisy.


\end{myitemize}

\myitem\cmt{General formulation of cartography \ra needs many
data}
\begin{myitemize}
\myitem\cmt{composition}Since the source of such an error is the
dependency of $\pilotfunest(\pilotsigmat)=\hat
 p(\hbm x(\pilotsigmat))$ on the estimated location $\hbm x(\pilotsigmat)$, one could think of
bypassing this dependence  by directly estimating  $\pilotfunest$ as a
general function of $\pilotsigmat$: 
\begin{alignat}{3}
\begin{aligned}  \label{eq:locasedDirect}
&\mathcal{Y}&\overset{\pilotfunest}{\xrightarrow{\hspace*{0.7cm}}} & \quad \rfield
\\
&\pilotsigmat&\xrightarrow{\hspace*{0.7cm}} & \quad \pilotfunest(\pilotsigmat).
\end{aligned}
\end{alignat}
\myitem\cmt{estimation}When pursuing an estimate of this general form, 
$\pilotfunest(\pilotsigmat)$ would not be confined to depend on
$\pilotsigmat$ only through the estimated location. However, finding
such an estimate given $\{(\pilotsigmat_n,\tilde{p}_n)\}_{n=1}^N$ by
searching over a generic class of functions such as an RKHS
\myitem\cmt{limitations}would be extremely challenging due the
so-called \emph{curse of
dimensionality}~\cite{bishop2006,cherkassky2007}. To intuitively
understand this phenomenon, note that the number of input variables of
function $\pilotfunest(\pilotsigmat)$ is $\sourcenum K$, typically in
the order of hundreds or thousands. Since learning a multivariate
function up to a reasonable accuracy generally requires that the number of
data points  be several orders of magnitude larger than the number
of input variables, this approach would need $N$ to be significantly
larger than $\sourcenum K$, and therefore prohibitively large.
\end{myitemize}%

\begin{figure}
\begin{subfigure}{4cm}
\includegraphics[width=4cm]{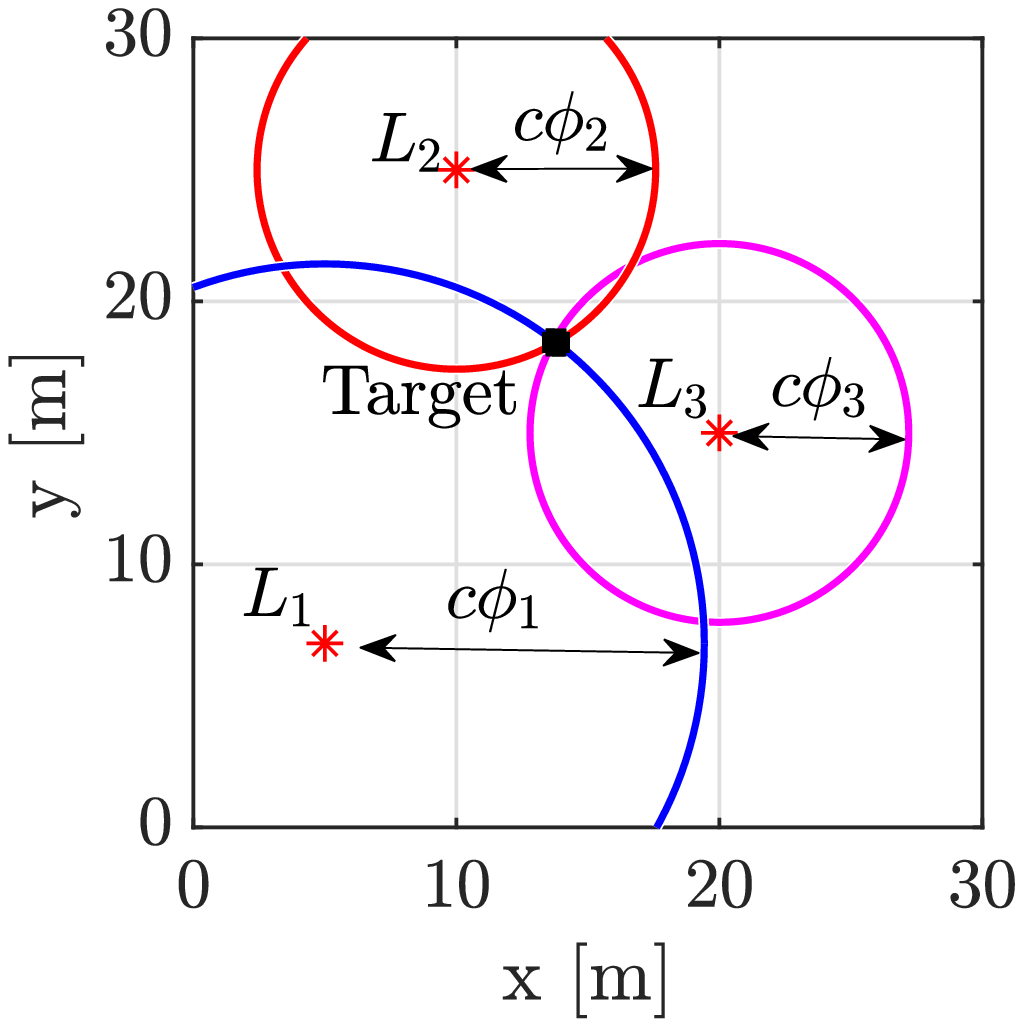}
\caption{}
\end{subfigure}
\begin{subfigure}{4cm}
\includegraphics[width=4cm]{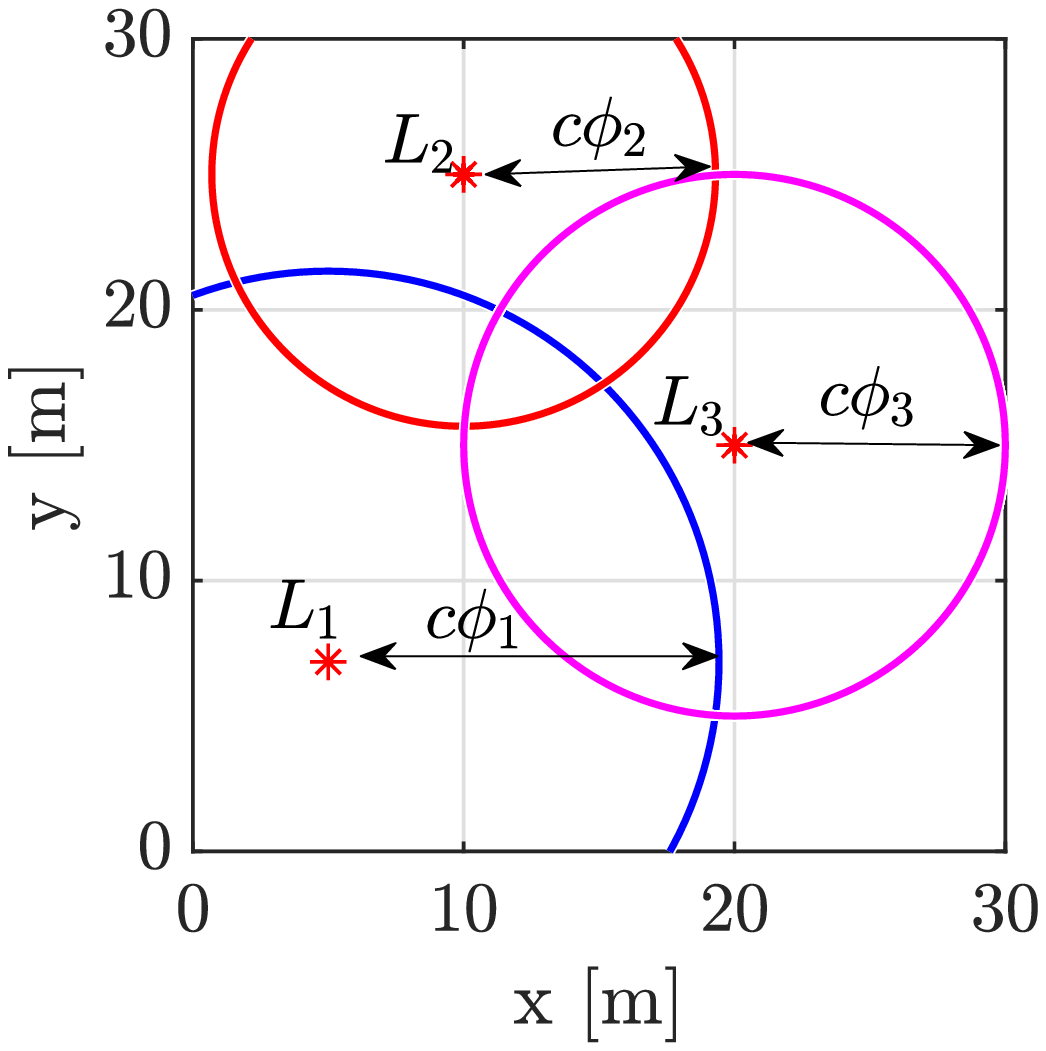}
\caption{}
\end{subfigure}
\caption{ Multi-lateration using ToA measurements with circles as possible sensor locations: (a) consistent ToA with the sought sensor location being the intersection of the circles (black square) and (b) inconsistent ToA measurements. The red stars represent the locations of the anchor base stations.} 
\label{f:ToAconsistency}
\end{figure}
\begin{myitemize}
\myitem\cmt{Tradeoff}To summarize,  the structure imposed by  \eqref{eq:locasedDirect} is too generic, whereas the one imposed by 
\eqref{eq:locasedComp} is too restrictive. To attain a sweet spot in this trade-off, it  is worth decomposing $\hbm x(\pilotsigmat)$ as detailed next. 
\myitem\cmt{composition}Recall that $\hbm x(\pilotsigmat)$ is the result of applying a localization algorithm to the pilot signals $\pilotsigmat$. For most existing algorithms, $\hbm x(\pilotsigmat)$ can be thought of as  the composition of two functions: a
function
$\featurevec:\mathcal{Y}\rightarrow \featureset\subset \rfield^{\featurenum}$ that
obtains $\featurenum$ features from $\pilotsigmat$, such as T(D)oA or DoA, and a function
$ \hbm l:\featureset\rightarrow \mathcal{X}$, that provides a location estimate
$ \hbm l(\featurevec)$ given a feature vector $\featurevec\in \featureset$. In this case,  $\pilotfunest(\pilotsigmat)$ can be decomposed~as:
\begin{alignat}{3}
\begin{aligned}  \label{eq:locasedFeat}
&\mathcal{Y} & \overset{\featurevec}{\xrightarrow{\hspace*{0.4cm}}} &  \quad {\featureset} & \overset{\hbm l}{\xrightarrow{\hspace*{0.4cm}}} &\quad \mathcal{X} & \overset{\hat p}{\xrightarrow{\hspace*{0.4cm}}}&  \quad \mathbb{R}
\\
&\pilotsigmat &\xrightarrow{\hspace*{0.4cm}}& \quad \featurevec(\pilotsigmat)& \xrightarrow{\hspace*{0.4cm}} & \quad \hbm l( \featurevec(\pilotsigmat))& \xrightarrow{\hspace*{0.4cm}}& \quad \hat p( \hbm l(\featurevec(\pilotsigmat))).
\end{aligned}
\end{alignat}

\myitem\cmt{limitations}Observe that the reason why the location
estimate $\hbm x(\pilotsigmat) = \hbm l(\featurevec(\pilotsigmat))$
is inaccurate in multipath environments is because the algorithm
that evaluates $\hbm l$ adopts a model where there is a certain
``agreement'' among features $\featurevec(\pilotsigmat)$.
\begin{myitemize}%
\myitem\cmt{Figure}To see this, consider Fig.~\ref{f:ToAconsistency}, which illustrates the task of estimating the location of a sensor in an area with $\sourcenum=3$ base stations. The features in $\featurevec \in \rfield^{\featurenum}$, with $\featurenum=\sourcenum=3$, used in this example are noiseless ToA features.
For each pilot signal, there is a circle centered at the base station and whose radius equals $c$ times the ToA, where $c$ is the speed of light.
\myitem\cmt{consistency in free space}Thus, when there is no multipath, the ToA features are accurate and the sensor to be located must lie in the intersection of the three circles, as shown in Fig.~\ref{f:ToAconsistency}a. Thus, the localization algorithm (embodied in $\hbm l$) just needs to return the location at which these circles intersect. 
\myitem\cmt{inconsistency in multipath}However, in multipath environments, the ToA features obtained from $\pilotsigmat$ do not generally equal the time it takes for an electromagnetic wave to propagate from the corresponding base station to the sensor.  As a result, the aforementioned circles will not generally intersect; see Fig.~\ref{f:ToAconsistency}b. In other words, the expected agreement among features is absent and, hence, the localization algorithm will return an inaccurate estimate of the position.
\end{myitemize}
\end{myitemize}

\myitem\cmt{Proposed: learn function of features}In view of these arguments, 
\begin{myitemize}%
\myitem\cmt{composition}the key idea in this paper is to pursue estimates  $\pilotfunest(\pilotsigmat)$ of the form:
\begin{alignat}{3}
\begin{aligned}  \label{eq:locFree}
&\mathcal{Y} & \overset{\phi}{\xrightarrow{\hspace*{0.7cm}}} &\quad \featureset & \overset{\featurefunest}{\xrightarrow{\hspace*{0.7cm}}}&  \quad \mathbb{R}
\\
&\pilotsigmat & \xrightarrow{\hspace*{0.7cm}} & \quad \featurevec(\pilotsigmat)& \xrightarrow{\hspace*{0.7cm}}& \quad \featurefunest(\featurevec(\pilotsigmat)).
\end{aligned}
\end{alignat}
\myitem\cmt{estimation}In this setting, the problem is find an estimate $\featurefunest(\featurevec)$ given $\{(\featurevec_n,\tilde
p_n)\}_{n=1}^N$, where $\featurevec_n := \featurevec(\pilotsigmat_n)$.
\myitem\cmt{benefits}By following  this approach, the estimated  map
$\hat p(\pilotsigmat)= \featurefunest(\featurevec(\pilotsigmat))$ does not involve a high number
 of inputs as in \eqref{eq:locasedDirect} and does not depend on the location
 estimate as in \eqref{eq:locasedComp}. For the latter reason, this approach will be
 referred to as \emph{LocF cartography}.  Since this approach
 does not need the agreement among entries of $\featurevec(\pilotsigmat)$
 illustrated in Fig.~\ref{f:ToAconsistency}b, it is expected to outperform
 traditional spectrum cartography methods when such an agreement is not
 present, as occurs in multipath environments.  

\end{myitemize}
\end{myitemize}
\subsection{Kernel-based Power Map Learning}
\label{sec:kernel}

\cmt{overview}This section applies kernel-based learning to provide an algorithm capable of learning the 
function $\featurefunest$ introduced in Sec.~\ref{sec:composition}.

\cmt{regression problem formulation}Given pairs
$\left\lbrace (\featurevec_{n},\tilde{p}_n)\right\rbrace _{n=1}^N$,
where $\featurevec_{n}:=\featurevec(\pilotsigmat_n)$, the problem can be informally stated as finding a function
$\featurefunest$ that satisfies two conditions:
\begin{myitemize}
\myitem\cmt{C1}C1) $\featurefunest$ fits
the data, that is,
$\featurefunest(\featurevec_{n})\approx \tilde{p}_n,~n=1,\ldots,N$; and
\myitem\cmt{C2}C2) $\featurefunest$ generalizes well to unseen data, i.e., if a new pair
$(\featurevec_{N+1},\tilde{p}_{N+1})$ is received, then
$\featurefunest(\featurevec_{N+1})\approx \tilde{p}_{N+1}$.
\end{myitemize}%
\cmt{kernel-based approach}%
\begin{myitemize}%
\myitem\cmt{motivation}A popular approach to solve the aforementioned
function learning problem is kernel-based learning, mainly due to its
simplicity, universality, and good performance~\cite{scholkopf2001}.
Furthermore, multiple works have demonstrated the merits of this framework
for spectrum cartography; see Sec.~\ref{sec:introduction}.

\myitem\cmt{background kernel-based learning}The first step when attempting to learn a function is to specify in which family of functions
$\featurefunest$ must be sought. In kernel-based learning, one seeks
$\featurefunest$ in a set known as a \emph{reproducing-kernel Hilbert
space} (RKHS), which is given by:
\begin{align} \label{eq:rkhs}
 \rkhs:=\left\lbrace \featurefun: \featurefun(\featurevec)=\sum_{i=1}^ \infty \alpha_i
 \kappa(\featurevec,\featurevec'_i),~\featurevec'_i \in \featureset,~ \alpha_i \in \mathbb{R}\right\rbrace,
\end{align}
where $\kappa:\featureset \times \featureset \rightarrow \mathbb{R}$ is a
symmetric and positive definite function known as \emph{reproducing
kernel}~\cite{scholkopf2001representer}. \arev{Although kernel methods can use any reproducing kernel}, a common choice is the
so-called Gaussian \emph{radial basis function} $
\kappa(\featurevec,\featurevec'):=\text{exp}\left[ - \Vert \featurevec-\featurevec'\Vert
^2/{(2\sigma^2)}\right] $, where $\sigma>0$ is a parameter selected by
the user. As any Hilbert space, $ \rkhs$ has an associated
inner product and norm. For an RKHS function
$\featurefun(\featurevec)=\sum_{i=1}^ \infty \alpha_i
\kappa(\featurevec,\featurevec'_i)$, 
\vspace{-0.2cm} the latter is
given by:
\begin{align}
\left \Vert \featurefun \right \Vert _{\rkhs}^{2} :=\sum_{i=1}^{\infty} \sum_{j=1}^{\infty}\alpha_i \alpha_j \kappa(\featurevec'_i,\featurevec'_j). \vspace{-0.2cm}
\end{align}
\vspace{-0.2cm}
\myitem\cmt{general kernel-based learning criterion}Kernel-based learning
typically solves a problem of the form:
\begin{align} \label{eq:pregr}
\featurefunest= \arg \min_{{\featurefun}\in \rkhs}\frac{1}{N}\sum_{n=1}^N \mathcal{L} \left(\tilde{p}_n, \featurevec_n, {\featurefun}(\featurevec_n)\right) +  \omega \left( \left \Vert {\featurefun} \right \Vert _{\rkhs} \right),
\end{align}
where
\begin{myitemize}
\myitem{}$\mathcal{L}$ is a loss function quantifying the deviation
between the observations $\{\tilde{p}_n\}_{n=1}^N$ and the predictions
 $\{\featurefun(\featurevec_n)\}_{n=1}^N$ returned by a candidate $\featurefun$;
\myitem{} and $\omega$ is an increasing function.
\end{myitemize}
The first term in \eqref{eq:pregr} promotes function estimates
satisfying C1. The second term promotes estimates satisfying C2 by
limiting overfitting. Intuitively, $\Vert \cdot \Vert _{\rkhs}$
captures a certain form of smoothness that limits the variability of
$\featurefun$. 
\myitem\cmt{krr criterion applied to our problem}
\begin{myitemize}

\myitem\cmt{KRR}\arev{Although there exist different candidate functions for $\mathcal{L}$  and $\omega$ in kernel-base learning,} typical choices are
 $\mathcal{L}(\tilde{p}_n, \featurevec_n, {\featurefun}(\featurevec_n))=
( \tilde{p}_n- {\featurefun}(\featurevec_n))^2 $ and $\omega( \Vert
{\featurefun} \Vert _{\rkhs})=
\regpar \Vert {\featurefun}\Vert _{\rkhs}^2$, where $\regpar
> 0$ is a regularization parameter that balances smoothness and
  goodness of fit. For this choice, $\featurefunest$ is
  termed \emph{kernel ridge regression}
  estimate~\cite[Ch. 4]{scholkopf2001}, and is the one used in our
  experiments for simplicity.
\end{myitemize}
\myitem\cmt{rep. theorem and estimator}The goal is therefore to
  solve \eqref{eq:pregr}. However, since $\rkhs$ is generally infinite
  dimensional, \eqref{eq:pregr} cannot be directly solved.
  Fortunately, one can invoke the \emph{representer theorem}~\cite{scholkopf2001representer},
  which states that the solution to \eqref{eq:pregr} is of the form:
\begin{align} \label{eq:estp}
  \featurefunest(\featurevec)=\sum_{n=1}^N \alpha_{n}\kappa(\featurevec,\featurevec_{n}),
  \end{align}
    for some $\{ \alpha_{n}\}_{n=1}^N$. 
  \begin{myitemize}%
\myitem{}Although the representer theorem does not provide  $\left\lbrace \alpha_{n}\right\rbrace _{n=1}^N$, these coefficients
  can be obtained by substituting \eqref{eq:estp}
  into \eqref{eq:pregr} and solving the resulting problem with respect
  to them.  Applying this procedure for kernel ridge
  regression results in the problem:
\begin{align} \label{eq:optalpha}
\hat{\bm{\alpha}}= \arg \min_{\bm{\alpha}} \frac{1}{N}\left \Vert \tilde{\bm{p}}-\bm{K}\bm{\alpha} \right \Vert ^2+ \regpar \bm{\alpha}^{\top}\bm{K}\bm{\alpha},
\end{align}
where
\begin{myitemize}
\myitem{}$\bm{\alpha}:=\left[\alpha_1,...,\alpha_N \right]^{\top} $, 
\myitem{}$\tilde{\bm{p}}:=\left[\tilde{p}_1,...,\tilde{p}_N \right]^{\top} $,
\myitem{}and $\bm K$ is a positive-definite ${N \times N}$ matrix whose $(n,n')$-th
entry is $\kappa(\featurevec_{n},\featurevec_{n'})$.
\end{myitemize}%
\myitem{}Problem~\eqref{eq:optalpha} can be readily solved in closed-form as $
\hbm{\alpha}= \left(\bm{K}+\regpar N \textbf{I}_{N}\right)^{-1} \tilde{\bm{p}}$.
\end{myitemize}%
\myitem \cmt{estimate at query points}The estimate $\featurefunest$ solving
\eqref{eq:pregr} for kernel ridge regression can be recovered by   substituting
the resulting  $\{ \alpha_{n}\}_{n=1}^N$ into \eqref{eq:estp}.  To obtain the predicted power
of the C2M at a query location $\bm {x}$ where the pilot signals are
given by $\pilotsigmat$, one just evaluates the LocF estimate
$\pilotfunest(\pilotsigmat)=\featurefunest(\featurevec(\pilotsigmat))$.

\end{myitemize}

\section{Location-Free Features}
\label{sec:features}

\cmt{Motivation feature engineering}
\begin{myitemize}
\myitem\cmt{LocB}As described in Sec.~\ref{sec:composition}, 
LocB cartography algorithms learn a function of the 
location estimate. In the machine learning terminology, the \emph{features} are
the spatial coordinates of the sensor locations.
\myitem\cmt{LocF}On the other hand, the features used by LocF cartography
are the entries of $\featurevec(\cdot)$. In principle,
$\featurevec(\pilotsigmat)$ could be set  to contain the same features
as the ones used by $\hbm l(\cdot)$; see
Sec.~\ref{sec:locfree}. However, it is generally preferable to use
features specifically tailored to LocF cartography. 
\end{myitemize}%
\cmt{overview}This section  accomplishes the design of these features in several steps.

\subsection{Feature Extraction}
\label{sec:featureExtr}

\cmt{Motivation}In Sec.\ref{sec:composition},  $\featurevec(\pilotsigmat)$ comprised $\featurenum$ features used by typical localization
algorithms, e.g. T(D)oA or DoA. The key observation is that, although
these features are appropriate for localization, a different set of
features may be preferable for LocF cartography.
\cmt{Overview}To come up with a
natural feature design, this section first reviews the features used
by typical localization algorithms (hence for LocB cartography) and
analyzes their limitations. Inspired by this analysis, a novel feature
extraction approach is proposed. To simplify the exposition, the
scenario where sensors are synchronized with the base stations is
presented first. A more practical setup, where this
synchronization is not required, will be considered next.



\subsubsection{Sensors are Synchronized with Base Stations}
\label{sec:sync}
\cmt{ToA}
\begin{myitemize}%
\myitem\cmt{model}The received pilot signal is generally modeled as:
\begin{align}
\label{eq:rxsamples}
y_{\sourceind,n}[k] := \pilotsig_\sourceind[k]\ast h_{\sourceind,n}[k] + \noisesig_{\sourceind,n}[k],
\end{align}
where
\begin{myitemize}
\myitem{}$\pilotsig_\sourceind[k]$ is the $k$-th sample of the $\sourceind$-th transmitted pilot signal,
\myitem{}$h_{\sourceind,n}[k] $ is the discrete-time channel impulse
response between the $\sourceind$-th base station and the sensor at
the $n$-th location, and
\myitem $\noisesig_{\sourceind,n}[k]$ is the noise term.
\end{myitemize}%
\cmt{channel}%
\begin{myitemize}%
\myitem\cmt{continuous time}The discrete-time impulse response $h_{\sourceind,n}[k]$ is obtained next from its analog counterpart $h_{\sourceind,n}(t)$, which 
follows the conventional multipath channel model with  $P_{\sourceind,n}$ components:
\begin{align}
\label{eq:continuoustimechannel}
h_{\sourceind,n}(t)=\sum_{p=1}^{P_{\sourceind,n}} \alpha_{\sourceind,n}^{(p)}\delta\left(t-t_{\sourceind,n}^{(p)}\right),
\end{align}
where
\begin{myitemize}%
\myitem{}$\delta(\cdot)$ is the Dirac delta distribution and
\myitem{}$\alpha_{\sourceind,n}^{(p)}\in \rfield$ and $t_{\sourceind,n}^{(p)}$ are respectively  the amplitude and delay of the $p$-th  
path. 
\end{myitemize}%
\myitem\cmt{equivalent discrete-time}After up-conversion to the
carrier frequency $f_c$, the pilot signal of the $\sourceind$-th base
station is transmitted and received by the sensor at the $n$-th measurement point, which
bandpass-filters with bandwidth $B$, down-converts, and samples at the Nyquist rate $T=1/B$. Therefore, the received noiseless samples are given by
$y_{\sourceind,n}[k]$ in \eqref{eq:rxsamples}, where~\cite{goldsmith,smith1997scientist}:
\begin{align}
\label{eq:discretetimechannel}
h_{\sourceind,n}[k]=\sum_{p=1}^{P_{\sourceind,n}} \alpha_{\sourceind,n}^{(p)}e^{-\sqrtmo2\pi f_c t_{\sourceind,n}^{(p)}} \sinc\left(k-\frac{t_{\sourceind,n}^{(p)}}{T}\right).
\end{align}
\end{myitemize}

\myitem\cmt{Conventional ToA}
\begin{myitemize}
\myitem\cmt{def}
In view of these expressions, one of the
most natural estimators for the ToA $\toa_{\sourceind,n}:=t_{\sourceind,n}^{(1)}$
is:
\begin{align}
\label{eq:toaestimate}
\hat \tau_{\sourceind,n}:=T\cdot\min\{k:|\hat h_{\sourceind,n}[k]|\geq \threshold\},
\end{align}
where
\begin{myitemize}
\myitem $\hat h_{\sourceind,n}[k]$ is an estimate of $
h_{\sourceind,n}[k]$ and
\myitem $\threshold$ is typically set as a function of the
signal-to-noise ratio~\cite{dardari2008threshold}.
\end{myitemize}

\begin{figure}
\centering
\begin{subfigure}{4cm}
\includegraphics[width=4cm]{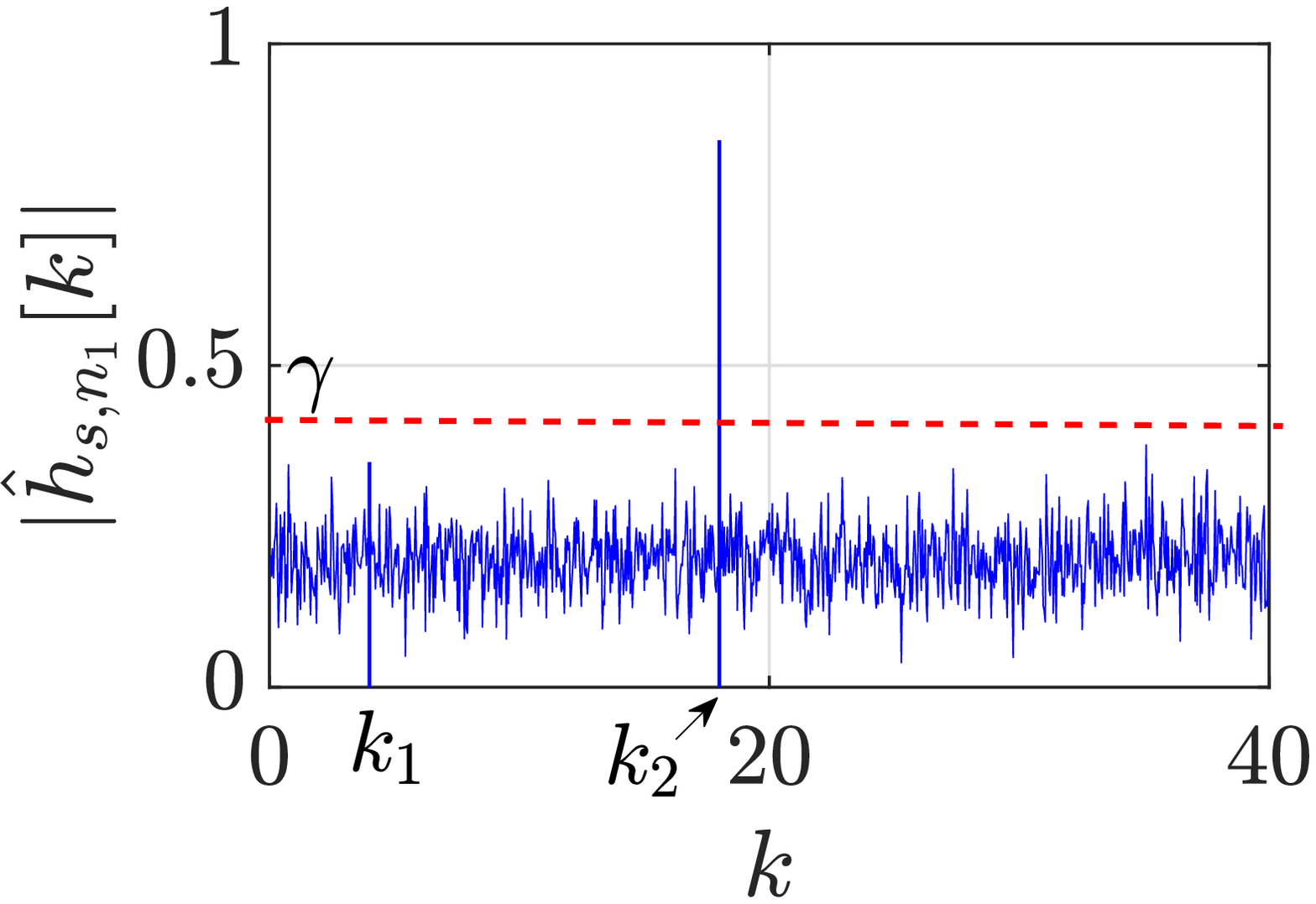}
\caption{}
\label{f:impulse2ToAunderthresh}
\end{subfigure}%
\begin{subfigure}{4cm}
\centering\includegraphics[width=4cm]{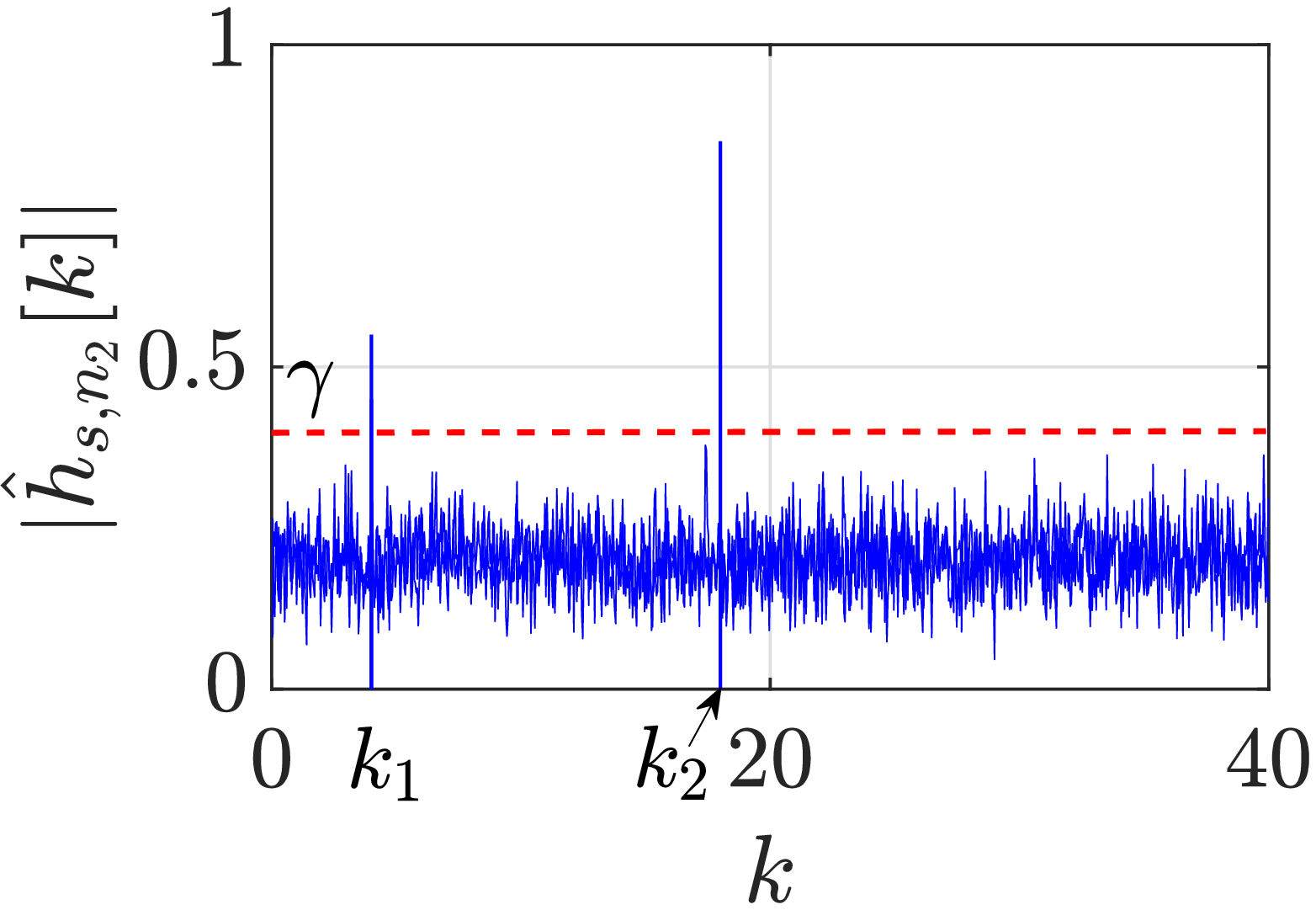}
\caption{}
\label{f:impulse2ToAoverthresh}
\end{subfigure}
\caption{Extraction of ToA from digital impulse response measured at two points that are spatially close. In (a), the ToA estimate is proportional to $k_2$; whereas in (b), the ToA estimate is proportional to $k_1$.}
\label{f:impulse2ToA}
\end{figure}

\myitem\cmt{ToA not smooth\ra example}
It will be argued next that such a ToA feature does not evolve
smoothly over space in presence of multipath, and therefore, this may negatively impact 
estimation performance, as occurs with LocB cartography; see
discussion about Fig.~\ref{f:locEst} in Sec.~\ref{sec:introduction}. 
For simplicity, assume that $\pilotsig_\sourceind[k]=\delta[k]$, where
$\delta[k]$ is the Kronecker delta.        In
this case, one can directly estimate $ h_{\sourceind,n}[k]$ as  $\hat
h_{\sourceind,n}[k] =y_{\sourceind,n}[k]  
= h_{\sourceind,n}[k]+ \noisesig_{\sourceind,n}[k]$, which is  a noisy 
version of $h_{\sourceind,n}[k]$. 
To see the impact  of multipath, 
consider
a simple example where the measurement points $\bm x_{n_1}$ and $\bm
x_{n_2}$ lie close to each other and the  channel impulse responses
are given by
$\hat
h_{\sourceind,n_1}[k]={\alpha}_{\sourceind,n_1}^{(1)} \delta[k-k_{\sourceind,n_1}^{(1)}]+
{\alpha}_{\sourceind,n_1}^{(2)} \delta[k-k_{\sourceind,n_1}^{(2)}]
+ \noisesig_{\sourceind,n_1}[k]$
 and 
$\hat
h_{\sourceind,n_2}[k]={\alpha}_{\sourceind,n_2}^{(1)} \delta[k-k_{\sourceind,n_2}^{(1)}]+
{\alpha}_{\sourceind,n_2}^{(2)} \delta[k-k_{\sourceind,n_2}^{(2)}]
+ \noisesig_{\sourceind,n_2}[k]$. Due to their spatial proximity, it
follows that:
\begin{subequations}
\label{eq:parameterproximity}
\begin{align}
{\alpha}_{\sourceind,n_1}^{(1)} \approx
{\alpha}_{\sourceind,n_2}^{(1)},\quad &
{\alpha}_{\sourceind,n_1}^{(2)} \approx
{\alpha}_{\sourceind,n_2}^{(2)},\\
k_{\sourceind,n_1}^{(1)}\approx
k_{\sourceind,n_2}^{(1)} \approx k_1,\quad &
k_{\sourceind,n_1}^{(2)} \approx  k_{\sourceind,n_2}^{(2)}\approx k_2,
\end{align}
\end{subequations}
for some $k_1$ and $k_2$. Assuming for simplicity that the effects of noise are negligible, if
$|{\alpha}_{\sourceind,n_1}^{(1)}| < \threshold <
|{\alpha}_{\sourceind,n_1}^{(2)}|
$
and
$\threshold<|{\alpha}_{\sourceind,n_2}^{(1)}|$, then the ToA estimates are:
\begin{align*}
\hat \toa_{n_1}&:=T\cdot\min\{k:|\hat
h_{\sourceind,n_1}[k]|\geq \threshold\} = T
k_{\sourceind,n_1}^{(2)} \approx T k_2,
\\\hat \toa_{n_2}&:=T\cdot\min\{k:|\hat
h_{\sourceind,n_2}[k]|\geq \threshold\} = T
k_{\sourceind,n_2}^{(1)}\approx T k_1.
\end{align*}
This scenario is illustrated in Fig.~\ref{f:impulse2ToA}. Despite how close their locations and observed impulse responses are,
the ToA estimates at locations  $\bm x_{n_1}$ and $\bm x_{n_2}$ can be quite different,
which establishes that the ToA estimate in \eqref{eq:toaestimate} is
not a smooth function of the spatial
location.
\end{myitemize}

\myitem\cmt{Proposed: CoM(impulse response)}
\begin{myitemize}
\myitem\cmt{def}Since this
non-smoothness negatively affects the performance of the proposed
LocF cartography estimator (and since the latter does not need
ToA estimates that are proportional to the distance, as occurs in
LocB cartography), a promising candidate for feature would
be the \emph{center of mass} (CoM) of the estimated impulse response:
\begin{align*}
\com_{\sourceind,n}:=\frac{\sum_{k=0}^{K-1}  |\hat h_{\sourceind,n}[k]|^2 k}{
\sum_{k=0}^{K-1} |\hat h_{\sourceind,n}[k]|^2},
\end{align*}
where $K$ is the number of samples. To see why such a feature evolves
smoothly over space, 
\myitem\cmt{example}suppose that the effects of noise are negligible
and note that this CoM feature applied to the channel impulse
responses in the previous example yields:
\begin{align*}
\com_{\sourceind,n_1}&=\frac{
k_{\sourceind,n_1}^{(1)}|\alpha_{\sourceind,n_1}^{(1)}|^2 +
k_{\sourceind,n_1}^{(2)}|\alpha_{\sourceind,n_1}^{(2)}|^2 }{
|\alpha_{\sourceind,n_1}^{(1)}|^2 + |\alpha_{\sourceind,n_1}^{(2)}|^2
},
\\
\com_{\sourceind,n_2}&=\frac{
k_{\sourceind,n_2}^{(1)}|\alpha_{\sourceind,n_2}^{(1)}|^2 +
k_{\sourceind,n_2}^{(2)}|\alpha_{\sourceind,n_2}^{(2)}|^2 }{
|\alpha_{\sourceind,n_2}^{(1)}|^2 + |\alpha_{\sourceind,n_2}^{(2)}|^2
}.
\end{align*}
From \eqref{eq:parameterproximity}, it follows that
$\com_{\sourceind,n_1}\approx \com_{\sourceind,n_2}$, which indicates
that the CoM is indeed a feature that evolves smoothly over
space, and
therefore preferable  for LocF cartography.
\myitem\cmt{feature vector}In this case, the feature vector at the
$n$-th sensor location becomes $\featurevec_n= 
[\com_{1,n},\ldots,\com_{\sourcenum,n}]^\top$.

\end{myitemize}
\end{myitemize}

\subsubsection{Sensors are not Synchronized with Base Stations}

\cmt{TDoA}Since synchronization requires more expensive equipment and
becomes challenging in multipath scenarios,  TDoA estimates are generally
preferred for localization. 
\begin{myitemize}%
\myitem\cmt{Typical TDoA estimator}TDoA estimates are typically
obtained
by extracting the lag corresponding to the maximum
cross-correlation of a pair of received pilot
signals~\cite{el2013low}. 
\begin{myitemize}%
\myitem\cmt{Cross-correlation}Assuming zero-mean, the cross-correlation between two pilot
signals received by the sensor at
the  $n$-th location is
defined as:
\begin{align}
\label{eq:crosscorrelation}
c\sensorsourcesourcenot{n}{\sourceind}{\sourceind'}[\lagind]:= \mathbb{E}\{ y_{\sourceind,n}[k]  y_{\sourceind',n}^{*}[k-\lagind] \} \quad \text{with} \quad \sourceind \neq \sourceind'.
\end{align}
\myitem With
$\pilotsig_\sourceind[k] = \pilotsig_{\sourceind'}[k]$ a white process
with power $\sigma_\pilotsig^2$ and uncorrelated with
$\noisesig_{\sourceind,n}[k]$ and $\noisesig_{\sourceind',n}[k]$, also
uncorrelated with each other, it can be easily seen that:
\begin{align*}
c\sensorsourcesourcenot{n}{\sourceind}{\sourceind'}[\lagind]= \sigma_{\pilotsig}^2\left(h_{\sourceind,n}[\lagind] \ast h_{\sourceind',n}^{*}[-\lagind]  \right).
\end{align*}
\myitem\cmt{arg max TDoA estimator}A common estimate
of the TDoA
${\tdoa}\sensorsourcesourcenot{n}{\sourceind}{\sourceind'}$ is  (see
e.g.~\cite{el2013low}):
\begin{align}
\label{eq:tdoafeat}
\hat{\tdoa}\sensorsourcesourcenot{n}{\sourceind}{\sourceind'}=T\cdot\arg \max_{\lagind}\{\left|\hat
c\sensorsourcesourcenot{n}{\sourceind}{\sourceind'}[\lagind]\right|\},
\end{align}
where $\hat
c\sensorsourcesourcenot{n}{\sourceind}{\sourceind'}[\lagind]$ is an estimate of $
c\sensorsourcesourcenot{n}{\sourceind}{\sourceind'}[\lagind]$.
To see the intuition behind this estimator, note that  $\hat
h_{\sourceind,n}[k]={\alpha}_{\sourceind,n}^{(1)} \delta[k-k_{\sourceind,n}^{(1)}]$
and $\hat
h_{\sourceind',n}[k]={\alpha}_{\sourceind',n}^{(1)} \delta[k-k_{\sourceind',n}^{(1)}]$
in a free-space channel with large bandwidth $B$.
This implies that: 
\begin{align*}
c\sensorsourcesourcenot{n}{\sourceind}{\sourceind'}[\lagind]&= \sigma_{\pilotsig}^2
{\alpha}_{\sourceind,n}^{(1)} \left({\alpha}_{\sourceind',n}^{(1)}\right)^* \delta\left[\lagind-\left(
k_{\sourceind,n}^{(1)} -k_{\sourceind',n}^{(1)}\right) \right]\\
&= \sigma_{\pilotsig}^2
{\alpha}_{\sourceind,n}^{(1)} \left({\alpha}_{\sourceind',n}^{(1)}\right)^* \delta\left[\lagind-
\tdoa\sensorsourcesourcenot{n}{\sourceind}{\sourceind'}/T \right],
\end{align*}
and therefore the lag of the maximum magnitude of
$c\sensorsourcesourcenot{n}{\sourceind}{\sourceind'}[\lagind]$
provides the TDoA in this simple scenario.
\end{myitemize}

\myitem\cmt{Limitations}Similar arguments to those used in
Sec.~\ref{sec:sync} to conclude that the 
 ToA estimates are not spatially smooth can also be invoked to reach
 the same conclusion  for TDoA. 
\myitem\cmt{Proposed: CoM(cross-correlation)}Likewise, following the
 same rationale as in Sec.~\ref{sec:sync}, this section proposes
 alleviating the aforementioned issue by adopting features of the form:
\begin{align} \label{eq:centerofmass}
\com\sensorsourcesourcenot{n}{\sourceind}{\sourceind'}
 :=\frac{\sum_{\lagind=-K+1}^{K-1} \vert
 c\sensorsourcesourcenot{n}{\sourceind}{\sourceind'}[\lagind] \vert ^2
 ~{\lagind}}{\sum_{\lagind=-K+1}^{K-1} \vert
 c\sensorsourcesourcenot{n}{\sourceind}{\sourceind'}[\lagind] \vert ^2 },
\end{align}
where $\com\sensorsourcesourcenot{n}{\sourceind}{\sourceind'}$ is  the  
CoM of the cross-correlation between the ${\sourceind}$-th and
${\sourceind'}$-th  pilot signals.  
\begin{myitemize}%
\myitem\cmt{strength of CoM}The proposed feature has three advantages:
i) it is smooth, as portrayed later in Sec.~\ref{sec:4loc_free_and_based}, ii) it does not require synchronization between the localization base stations and the sensors, and iii) it does not require the knowledge of the impulse responses.
\myitem\cmt{feature vector}With this choice,  the feature vector at
the $n$-th measurement  location becomes: \begin{align}
\label{eq:comfeatvecnosync}
\begin{split}
\featurevec_n =&
[\com\sensorsourcesourcenot{n}{1}{2},
\com\sensorsourcesourcenot{n}{1}{3},
\ldots,
\com\sensorsourcesourcenot{n}{1}{\sourcenum}, \\ &
\com\sensorsourcesourcenot{n}{2}{3}, \ldots,\com\sensorsourcesourcenot{n}{\sourcenum-1}{\sourcenum}]^\top.
\end{split}
\end{align}
\end{myitemize}%
\end{myitemize}%
\subsection{Cartography from a Reduced Set of Features}
\label{sec:reducedFeat}
\cmt{motivation}As argued earlier in
Sec.~\ref{sec:composition}, learning becomes difficult when the number
of input features $\featurenum$ is high.
\cmt{overview}This section develops a scheme to reduce this number of
features to improve estimation performance in LocF cartography.

\begin{myitemize}
\myitem\begin{myitemize}
\myitem\cmt{LocB}As stated in the previous section, in LocB cartography, the feature
vectors correspond to the coordinates of the estimated
location. Application of the localization algorithm represented by the
function $\hbm l$ in \eqref{eq:locasedFeat} naturally reduces
dimensionality from the original $\featurenum$ features to just 2 or
3.
\myitem\cmt{LocF}On the other hand,  in the case of
LocF cartography, a larger number $N$ of measurements to learn
$\featurefunest$ in \eqref{eq:locFree} may be necessary to attain a target accuracy
if $\featurenum$ is large. This observation calls for a dimensionality
reduction step that condenses the information of the feature vectors
$\{ \featurevec_{n}\}_{n=1}^N\subset \rfield^\featurenum$ into vectors
$\{ \featurecoordinatevec_{n}\}_{n=1}^N\subset \rfield^{\newfeaturenum}$ of a
reduced size $\newfeaturenum$. 
\end{myitemize}
\begin{figure}[t] 
\centering\includegraphics[width=0.55\columnwidth]{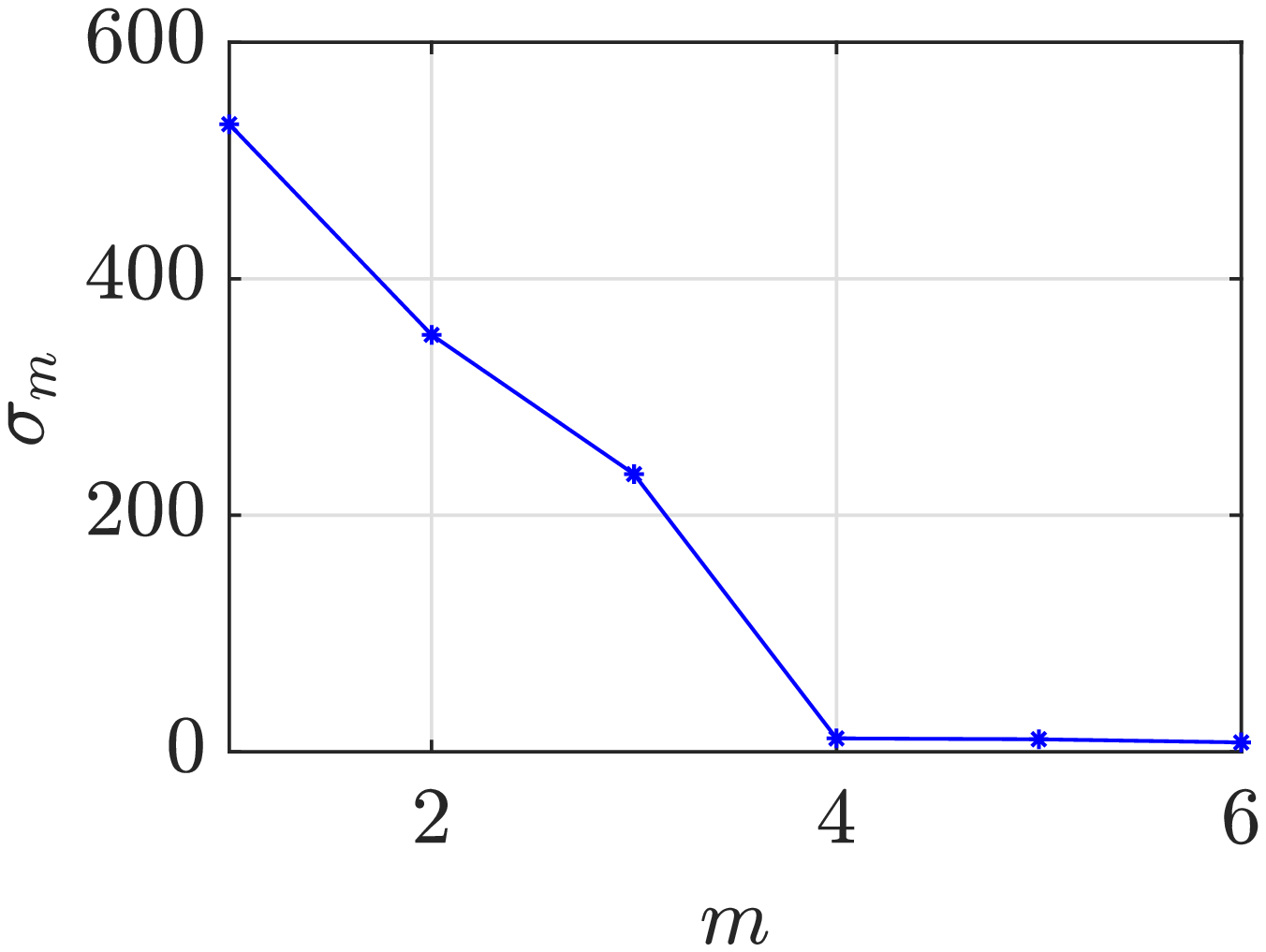} 
\caption{Singular values $\sigma_{1}\geq \sigma_{2}\geq\ldots\geq\sigma_{\featurenum}\geq 0$ of $\featuremat$ in
non-increasing order for a multipath environment with $\sourcenum=4$ transmitters.} 
\label{f:svdFeat}
\end{figure}
\myitem\cmt{preserve most info}Intuitively, $\newfeaturenum$ 
should be the minimum number that preserves most  information  while eliminating most of the noise in
$\{ \featurevec_{n}\}_{n=1}^N$. Even if some
information is lost, the reduction in the error entailed by the fact
that  the function to be estimated has fewer input arguments  may pay
off in practice.

\myitem\cmt{close to hyperplane}In the cases where the feature vectors 
$\{ \featurevec_{n}\}_{n=1}^N$ lie close to a low-dimensional
subspace, the coordinates of these vectors with respect to a basis for
such a subspace may constitute a  suitable reduced set of features.
To see this, it is instructive to start by considering the scenario of TDoA features.  
\begin{myitemize}
\myitem\cmt{TDoA}Suppose, for simplicity, that the effects of noise are
negligible, so that the TDoA estimates
$\hat{\tdoa}\sensorsourcesourcenot{n}{\sourceind}{\sourceind'}$
approximately equal the true TDoAs
${\tdoa}\sensorsourcesourcenot{n}{\sourceind}{\sourceind'}$.
Then, the rows of 
$\featuremat:=[\featurevec_1,\ldots,\featurevec_N]$
are of the form $\bm \tdoa_{{\sourceind},{\sourceind'}}:=[ \tdoa\sensorsourcesourcenot{1}{\sourceind}{\sourceind'},
\tdoa\sensorsourcesourcenot{2}{\sourceind}{\sourceind'},
\ldots,
\tdoa\sensorsourcesourcenot{N}{\sourceind}{\sourceind'} 
]^\top$. If
$\bm\toa_{\sourceind}:=[\toa_{\sourceind,1},\ldots,\toa_{\sourceind,N}]^\top$
collects the ToA from the $\sourceind$-th base station to all sensor
locations, then it clearly holds that 
$\bm\tdoa_{\sourceind,\sourceind'}= \toavec_{\sourceind}
- \toavec_{\sourceind'}$. Consequently, 
 $\bm \tdoa_{1,\sourceind}-
\bm \tdoa_{1,\sourceind'}
=
\toavec_{1}
- \toavec_{\sourceind} -
(
\toavec_{1}
- \toavec_{\sourceind'} 
)
=
 \toavec_{\sourceind'} 
- \toavec_{\sourceind}
=
\bm \tdoa_{\sourceind',\sourceind}
$, which implies that all rows of $\featuremat$ are linear combinations of the 
$\sourcenum-1$ rows
$\{\bm \tdoa_{1,\sourceind}\}_{\sourceind=2}^\sourcenum$.
Thus, the rank of $\featuremat$ is at most
$\sourcenum-1$ or, equivalently, the vectors
$\{ \featurevec_{n}\}_{n=1}^N$ lie in a subspace of dimension
$\sourcenum-1$. When effects of noise are noticeable, one would expect
that the vectors $\{ \featurevec_{n}\}_{n=1}^N$  lie  \emph{close} to
a subspace  of dimension $\sourcenum-1$. 


\myitem\cmt{CoM}
\begin{myitemize}
\myitem\cmt{Theoretical}Similarly, one can expect that when the
entries of  the
vectors $\{ \featurevec_{n}\}_{n=1}^N$ are  given by
 \eqref{eq:centerofmass}, these vectors  also lie close to a low-dimensional
subspace since CoM features are proportional to the TDoAs in absence of
multipath; see Sec.~\ref{sec:featureExtr}.
\myitem\cmt{Empirical}This phenomenon can be illustrated through
simulation (see Sec.~\ref{sec:numericalTest} for more details).
Fig.~\ref{f:svdFeat} depicts the singular values
$\sigma_{1}\geq \sigma_{2}\geq\ldots\geq\sigma_{\featurenum}\geq 0$ of
$\featuremat$ in non-increasing order for a multipath environment
described in Sec.~\ref{sec:numericalTest} with $\sourcenum=4$. As
expected, roughly $\newfeaturenum=\sourcenum-1=3$ directions capture
almost all the energy of the rows of $\featuremat$.
\end{myitemize}%
\end{myitemize}%

\end{myitemize}%
\cmt{PCA}\begin{myitemize}\myitem\cmt{motivation}When a set of random
vectors lie 
close to a subspace, an appealing approach for dimensionality
reduction is principal component analysis
(PCA)~\cite[Ch. 12]{bishop2006}, which obtains the reduced feature
vectors by projecting the input data vectors onto the
subspace that preserves most of the
energy.
\myitem\cmt{overview}Since in this paper no probabilistic assumptions have
been introduced on  $\{ \featurevec_{n}\}_{n=1}^N$, the typical
formulation of PCA is not directly applicable. However, as detailed
next, it is not difficult to extend this idea to the fully deterministic
scenario, which furthermore provides intuition.
\myitem\cmt{description}%

\begin{myitemize}%
\myitem\cmt{zero-mean}Assume w.l.o.g. a centered
set of feature vectors, i.e., $({1/N})\sum_{n=1}^N\featurevec_n=\bm
0$. If not centered, just subtract the mean by replacing $\featuremat$ with
$\featuremat -(1/N)\featuremat\bm 1\bm 1^\top$.
\myitem\cmt{svd}The subspace that captures most of
the energy of the observations can be determined using the singular
value decomposition (SVD) of  $\featuremat$, which for $\featurenum<N$
is given by:
\begin{align}
\label{eq:featurematsvd}
 \featuremat = \begin{bmatrix}
   \bm U_1 & \bm U_2
 \end{bmatrix}
\left[\begin{array}{@{}c|c@{}}
  \begin{matrix}
  \bm \Sigma_1 & \bm 0   \\\
  \bm 0        & \bm \Sigma_2
  \end{matrix}
\end{array} 
  \begin{matrix}
\  \bm 0   \\\
  \bm 0       
  \end{matrix} 
\right]
 \begin{bmatrix}
   \bm V_1^\top \\\
   \bm V_2^\top 
 \end{bmatrix},
\end{align}
where
\begin{myitemize}%
\myitem\cmt{svals}%
$\bm \Sigma_1:=\diag{\sigma_{1},\ldots,\sigma_{\newfeaturenum}}$
contains the $\newfeaturenum$ largest singular values of $\featuremat$,
$\bm \Sigma_2:=\diag{\sigma_{\newfeaturenum+1},\ldots,\sigma_{\featurenum}}$
contains the $\featurenum-\newfeaturenum$ smallest, 
\myitem\cmt{svecs}and the columns of  $\bm U := \begin{bmatrix}
   \bm U_1 & \bm U_2 \end{bmatrix}$ (respectively  $\bm V:= [
   \bm V_1, 
   \bm V_2]$) are the left (right) singular vectors of
 $\featuremat$.
 \end{myitemize}
\myitem\cmt{rotation preserves info}Clearly, if the data vectors $\{\featurevec_n\}_{n=1}^N$ are
   multiplied by the orthogonal matrix $\bm U^\top$, the resulting
   vectors $\{\rotatedfeaturevec_n\}_{n=1}^N$, with
   $\rotatedfeaturevec_n:=\bm U^\top \featurevec_n$, contain the same
   information. Thus, one can replace $\featuremat$ with
   $\rotatedfeaturemat:=\bm U^\top \featuremat$.

\myitem\cmt{keep entries with highest energy}
By applying this transformation, which can be thought of as a
generalized rotation, most of the energy of
$\rotatedfeaturemat$ is concentrated in its first $\newfeaturenum$
rows. To see this, note that the energy of the first $\newfeaturenum$
rows of $\rotatedfeaturemat$  is given by:
\begin{align*}
\vert\vert \bm
U_1^\top \featuremat \vert\vert_F^2&=\vert\vert \bm \Sigma_1  \bm
V_1^\top \vert\vert_F^2=\trace \left( \bm \Sigma_1  \bm V_1^\top \bm
V_1 \bm \Sigma_1^\top \right)\\&=\trace \left( \bm\Sigma_1 \bm\Sigma_1^\top \right)=\vert\vert \bm \Sigma_1 \vert\vert_F^2
=\sum_{\featureind=1}^\newfeaturenum\sigma_\featureind^2,
\end{align*}
whereas the energy of the last $\featurenum-\newfeaturenum$ rows of
$\rotatedfeaturemat$ is given by:
\begin{align*}
\vert\vert \bm
U_2^\top \featuremat \vert\vert_F^2=\vert\vert \bm \Sigma_2 \vert\vert_F^2
=\sum_{\featureind=\newfeaturenum+1}^\featurenum\sigma_\featureind^2.
\end{align*}
When $\newfeaturenum = \sourcenum-1$, since the rows of $\featuremat$ lie
approximately in a subspace of 
dimension $\newfeaturenum$, it follows that $\sigma_\featureind\approx 0$
for $\featureind>\newfeaturenum$. Therefore 
$\sum_{\featureind=1}^\newfeaturenum\sigma_\featureind^2\gg\sum_{\featureind=\newfeaturenum+1}^\featurenum\sigma_\featureind^2$
and, hence,  $
\vert\vert \bm U_1^\top \featuremat \vert\vert_F^2
\gg
\vert\vert \bm U_2^\top \featuremat \vert\vert_F^2
$. Equivalently, most of the energy of the vectors
$\{\rotatedfeaturevec_n\}_{n=1}^N$ is concentrated in their first
$\newfeaturenum$ entries.  This observation suggests using the first
$\newfeaturenum$ entries of the vectors
$\{\rotatedfeaturevec_n\}_{n=1}^N$ as features, while discarding the
rest. That is, the reduced dimensionality feature vectors will be given by
$\{\featurecoordinatevec_n\}_{n=1}^N$, where
$\featurecoordinatevec_n:=\bm U_1^\top \featurevec_n$. Note that  $\featurecoordinatevec_n$ is
just the vector of coordinates of $\featurevec_n$ with respect to the
basis composed of the columns of $\bm U_1$.

The number
$\newfeaturenum$ of entries of the new feature vectors
$\{\featurecoordinatevec_n\}_{n=1}^N$ may be potentially much smaller
than $\featurenum$ and can 
therefore boost estimation performance meaningfully. For instance, when
$\{\featurevec_n\}_{n=1}^N$ are given by 
\eqref{eq:comfeatvecnosync}, this reduction is from 
$\featurenum= \sourcenum(\sourcenum-1)/2$ features to
$\newfeaturenum=\sourcenum-1$ features.

\end{myitemize}
\myitem\cmt{choose alt. dimension}In scenarios of very strong
multipath, the rows of $\featuremat$ may not lie close to any subspace
of dimension $\sourcenum-1$. In those cases, it may be worth choosing
a value of $\newfeaturenum$ greater than $\sourcenum-1$. A possibility
is to specify a fraction $\retainedvar\in[0,1]$ of the energy of $\featuremat$
that must be kept in $\newfeaturemat:=\bm U_1^\top \featuremat$, and choose $\newfeaturenum$ to
be the smallest integer that guarantees this condition, that is:
\begin{align} \label{eq:retainedvar}
\newfeaturenum=\min\left\{\newfeaturenum': \frac{\sum_{\newfeatureind=1}^{\newfeaturenum'} \sigma_{\newfeatureind}^2
}{\sum_{\featureind=1}^\featurenum \sigma_{\featureind}^2}
\geq
\retainedvar\right\}.
\end{align}
\cmt{summary}To summarize, the problem of LocF cartography with the
technique for reducing the set of features introduced in this section
is as follows. Given the original set of measurements
$\{ \featurevec_{n}\}_{n=1}^N\subset \rfield^\featurenum$, one must
form the matrix $\featuremat$, compute $\bm U_1$ from the SVD
in \eqref{eq:featurematsvd}, and obtain the reduced features
$\{ \featurecoordinatevec_{n}\}_{n=1}^N\subset \rfield^{\newfeaturenum}$
where $\featurecoordinatevec_{n} = \bm
U_1^{\top}  \featurevec_{n}$. Then, the function $\featurefunest$
is obtained form the pairs $\{ (\featurecoordinatevec_{n},\tilde
p_n)\}_{n=1}^N$ using the approach in Sec.~\ref{sec:kernel}. To
evaluate the resulting map at a query location where the received pilot signals
are given by $\pilotsigmat$, one must simply obtain
$\featurefunest(\bm U_1^{\top}\featurevec(\bm Y))$.

\end{myitemize}
\subsection{Dealing with Missing Features}
\label{sec:missing}
\cmt{Motivation}
\begin{myitemize}
\myitem\cmt{missing pilots\ra missing features}Due to
propagation effects, the signal-to-noise ratio of some of the received pilot signals
may be too low for feature extraction. In this case, the features
associated with those pilot signals may be unreliable or simply
unavailable. 
\myitem\cmt{overview}This section develops techniques to cope with such missing features.
\end{myitemize}

\cmt{Notation}
\begin{myitemize}
\myitem\cmt{index set}Let
$\Omega \subset \{1,\ldots,\featurenum\}\times \{1,\ldots, N\}$ be
such that $(\featureind,n) \in \Omega$ iff the $\featureind$-th
feature is available at the $n$-th measurement location and define
\myitem\cmt{feat. mat}the ``incomplete'' feature matrix~$\missfeaturemat ~\in  \left( \rfield \cup \{ \missingfeat \} \right) ^{\featurenum \times N}$ as:
\begin{align}
 (\missfeaturemat)_{\featureind, n}=\begin{cases}
              (\featurevec_n)_{\featureind}+ \varsigma_{\featureind, n} & \text{if} ~
              (\featureind,n) \in \Omega \\ \missingfeat
              & \text{otherwise}, \end{cases}
\end{align}
where $\varsigma_{\featureind, n}$ explicitly models error in the feature
extraction and the symbol $\missingfeat$   represents that the
corresponding feature is missing.
\end{myitemize}%
\cmt{matrix completion}%
\begin{myitemize}%
\myitem\cmt{motivation}Since the matrix $\missfeaturemat$  contains
missing features, the  LocF cartography scheme presented so far is
not directly applicable. The
missing features must be filled first.
\myitem\cmt{goal}Hence, the goal is, given $\missfeaturemat$,
                  find $ \featuremat \in  \rfield
                  ^{\featurenum \times N}$
that agrees with $\missfeaturemat$ on $\Omega$. 
\myitem\cmt{rank minimization}%
\begin{myitemize}%
\myitem\cmt{motivation}A popular approach to address such a
                  matrix completion task is 
\myitem\cmt{opt. problem}via rank minimization~\cite{fazel2002matrix}:
\begin{align}  \label{eq:lowrank}
\begin{split}
& \underset{\featuremat}{ \text{minimize}}\quad \text{rank}\left( \featuremat \right) \\ & \text{subject to} \quad  \mathcal{P}_{\Omega}(\featuremat)=\mathcal{P}_{\Omega}(\missfeaturemat),
\end{split}
\end{align}
where 
\begin{alignat*}{3}
\begin{aligned} 
&\mathcal{P}_{\Omega}:  & \left( \rfield \cup \{ \missingfeat \} \right)^{\featurenum \times N}   & \longrightarrow  \rfield^{\featurenum \times N} \\
& & \missfeaturemat  & \longmapsto \mathcal{P}_{\Omega} (\missfeaturemat),
\end{aligned}
\end{alignat*}
with 
\begin{align*}
\left( \mathcal{P}_{\Omega}(\missfeaturemat) \right)_{\featureind, n}= \begin{cases}
             (\missfeaturemat)_{\featureind, n}
                  & \text{if} ~ (\featureind,n) \in \Omega \\ 0
                  &  \text{if} ~ (\featureind,n) \notin \Omega. \end{cases}
\end{align*} 

\myitem\cmt{solver \ra convex relaxation}Although this problem is
                  non-convex, efficient solvers exist based on convex
                  relaxation~\cite{candes2009exact,recht2010guaranteed}.
\myitem\cmt{recovery guaranties}A legitimate question would be what is
                  the minimum number of available features required to
                  recover a reasonable reconstruction of
                  $\featuremat$. As a guideline, a result
                  in~\cite{candes2010power} establishes that, under
                  certain conditions, the minimum number of available
                  features to recover
                  $\featuremat\in\rfield^{\featurenum\times N}$ is                   $\mathcal{O}\left(\tilde{N}\rank(\featuremat)\log(\tilde{N})\right)$
                  where $\tilde{N}=\max(\featurenum,N)$.
\end{myitemize}%
\myitem\cmt{Manifold opt}
\begin{myitemize}%

\myitem\cmt{Motivation}Although the aforementioned rank minimization
approach could, in principle, be used, it suffers from two
limitations. First, it does not exploit the prior information 
that 
$\featuremat$ can be well approximated by a matrix of rank
$\desiredrank$, where $\desiredrank$ is typically $\sourcenum-1$; see
Fig.~\ref{f:svdFeat}. Second, the constraint in \eqref{eq:lowrank}
would render the reconstructed matrix sensitive to the noise
$\{\varsigma_{\featureind,n}\}_{\featureind,n}$ present in
$\missfeaturemat$. Thus, an appealing alternative to  \eqref{eq:lowrank}
 would be: 
\myitem\cmt{opt. problem}
\begin{align}  \label{eq:manifoldopt}
\begin{split}
\completedfeaturemat:=& \underset{\featuremat}{ \text{argmin}}\quad \frac{1}{2} \vert \vert \mathcal{P}_{\Omega}(\featuremat)-\mathcal{P}_{\Omega}(\missfeaturemat
) \vert \vert_F^2 \\ & \text{subject
to} \quad \featuremat \in \mathcal{M}_\desiredrank,
\end{split}
\end{align} 
where $\mathcal{M}_\desiredrank:=\{ \featuremat \in \rfield  ^{\featurenum \times N}: \text{rank}(\featuremat)=\desiredrank \}$ is the smooth manifold of $\desiredrank$-rank $\featurenum \times N$ matrices.

\myitem\cmt{solvers}There exist algorithms to
find local minima of the non-convex problem~\eqref{eq:manifoldopt}. 
\begin{myitemize}%
\myitem\cmt{LRGeomCG}One example based on manifold
optimization~\cite{absil2009optimization} is the linear retraction-based geometric conjugate
gradient (LRGeomCG) method from~\cite{vandereycken2013low}.
\myitem\cmt{SVP}A less  computationally expensive alternative
is the singular value projection (SVP) method
in~\cite{jain2010guaranteed}, which is based on the traditional projected
subgradient descent method.

\end{myitemize}

\end{myitemize}
\end{myitemize}
\cmt{Dimensionality reduction}After
solving \eqref{eq:manifoldopt}, all the columns of $\completedfeaturemat:=[\completedfeaturevec_1,\ldots,\completedfeaturevec_N]$ clearly lie in a subspace of
dimension $\desiredrank$. From the arguments in
Sec.~\ref{sec:reducedFeat}, learning the map can be improved by
suppressing this redundancy. 
\begin{myitemize}
\myitem\cmt{PCA}To this end, one could use the technique  in
Sec.~\ref{sec:reducedFeat}, which would obtain the reduced-dimensionality
feature vectors as follows:
\begin{align} \label{eq:projectFeat}
\featurecoordinatemat:=[\featurecoordinatevec_1,\ldots,\featurecoordinatevec_N]=
\mathring{\bm U}_1^{\top} \completedfeaturemat.
\end{align}
Here, the columns of $\mathring{\bm U}_1$ are the  left singular vectors corresponding to the $\desiredrank$
 largest singular values of 
 $\completedfeaturemat$.
\myitem\cmt{Gram-Schmidt}Nevertheless, 
since $\completedfeaturemat$ has rank $\desiredrank$, it is not
necessary to obtain  $\mathring{\bm U}_1$  by means of an SVD. Namely, the
columns of $\mathring{\bm U}_1$ can be directly obtained by orthonormalizing the
first $\desiredrank$ linearly independent columns of
$\completedfeaturemat$, e.g. through Gram-Schmidt.

\cmt{summary}To sum up, to estimate a map using the proposed LocF
cartography in presence of missing features is as follows. First,
matrix $\missfeaturemat$ is formed with the available features. Then,
the completed matrix $\completedfeaturemat$ is obtained using
LRGeomCG or SVP. Next, $\mathring{\bm U}_1$ is obtained through Gram-Schmidt
over this completed matrix. Finally, one learns $\featurefunest$ from
$\{(\featurecoordinatevec_n,\tilde p_n)\}_{n=1}^N$, where
$\featurecoordinatevec_n$ is the $n$-th column of
$\featurecoordinatemat$ in \eqref{eq:projectFeat}, using the approach
in Sec.~\ref{sec:kernel}.

\end{myitemize}
\cmt{Evaluating the map with missing features}
\begin{myitemize}
\myitem\cmt{motivation}To  evaluate the estimated map at a
test location, one would require in principle the feature vector
$\featurevec\in \rfield^\featurenum$ at that location or,
alternatively, its reduced-dimensionality version
$\featurecoordinatevec\in \rfield^\desiredrank$.  However, due to
the phenomena described earlier, only some of the features of
$\featurevec$ may be available, which can be collected in the vector
 $\testmissfeaturevec \in \left( \rfield \cup \{ \missingfeat \} \right)^\featurenum$.
\myitem\cmt{problem}The problem now is to find
the 
  reduced-dimensionality feature vector $\featurecoordinatevec$  given
$\testmissfeaturevec$.

\myitem\cmt{prior knowledge}Since the columns of $\completedfeaturemat$ lie in
an $\newfeaturenum$-dimensional subspace for which the columns of
$\mathring{\bm U}_1$ form an orthonormal basis, it is reasonable to say that the
feature vector at the testing point
$\featurevec \in \rfield^{\featurenum}$ also lies in that subspace,
meaning that this vector can be written as $\featurevec=\mathring{\bm U}_1 \featurecoordinatevec$ for some $\featurecoordinatevec$.
\myitem\cmt{Procedure\ra depends on num. obs. feat.}The procedure to
recover $\featurecoordinatevec$ depends on whether
$\testmissfeaturevec$ contains enough observed
features. Let $\Omega'\subset\{1,\ldots, \featurenum\}$ be such that
the $\featureind\in\Omega'$ iff the $\featureind$-th feature is
available in $\testmissfeaturevec$.  
\begin{myitemize}%
\myitem\cmt{sufficient num. feat.}If $\obsfeaturenum:=|\Omega'| \geq \desiredrank$,
\begin{myitemize}
\myitem\cmt{regularized LS}one can think of finding
$\featurecoordinatevec$  using the well-known regularized least
squares (RLS) method as:
\begin{align} \label{eq:evalCoeff}
\begin{split}
\hat \featurecoordinatevec =  \arg \min_{\featurecoordinatevec} & \left \Vert   \mathcal{P}_{\Omega'}(\testmissfeaturevec ) - \mathcal{P}_{\Omega'}(\mathring{\bm U}_1 \featurecoordinatevec)  \right \Vert ^2 \\ &+ \evalregpar (\featurecoordinatevec-\featurecoordinatevec_{\text{avg}})^{\top}\bm C^{-1}(\featurecoordinatevec-\featurecoordinatevec_{\text{avg}}),
\end{split}
\end{align}
where
\begin{myitemize}
\myitem\ 
\begin{alignat*}{3}
\begin{aligned} 
&\mathcal{P}_{\Omega'}:  & \left( \rfield \cup \{ \missingfeat \} \right)^{ \featurenum}   & \longrightarrow  \rfield^{ \featurenum} \\
& & \testmissfeaturevec & \longmapsto
\featurevec,~ (\featurevec)_{\featureind}=
\begin{cases}
(\testmissfeaturevec)_{\featureind}  & \text{if}~ \featureind \in \Omega'  \\ ~ 0
                  &  \text{if}~ \featureind \notin \Omega', \end{cases}
\end{aligned}
\end{alignat*}
\myitem{}$\evalregpar>0$ is a regularization parameter,
\myitem{}$\featurecoordinatevec_{\text{avg}}$ and  $\bm C \in \rfield^{\newfeaturenum\times \newfeaturenum}$
                  are respectively the sample mean vector and  covariance matrix of the coordinates
                  of the completed features in the
                  traning phase, that is,
$\featurecoordinatevec_{\text{avg}} = (1/N) \newfeaturemat\bm 1$ and $\bm C =
                  (1/N)( \newfeaturemat
                  - \featurecoordinatevec_{\text{avg}}\bm 1^\top)
                  ( \newfeaturemat
                  - \featurecoordinatevec_{\text{avg}}\bm 1^\top)^\top$.
\end{myitemize}%
\myitem\cmt{closed-form sol}To solve Problem~\eqref{eq:evalCoeff}, let
the elements of $\Omega'$ be denoted as $\Omega'
:=\{\featureind_1,\ldots,\featureind_{\obsfeaturenum}\}$. Then:
\begin{align} \label{eq:cfevalCoeff}
\begin{split}
\hat{\featurecoordinatevec}= & \left( \mathring{\bm U}_1^{\top} \bm S^{\top} \bm S \mathring{\bm U}_1 + \evalregpar  \bm C^{-1}  \right)^{-1} \\ & \left( \mathring{\bm U}_1^{\top} \bm S^{\top}\bm S \mathcal{P}_{\Omega'}(\testmissfeaturevec )  +  \evalregpar  \bm C^{-1} \featurecoordinatevec_{\text{avg}} \right), 
\end{split}
\end{align}
where $\bm S\in\{0,1\}^{\obsfeaturenum\times\featurenum}$ is a row
selection matrix with all entries equal to zero except for the entries
$(1,\featureind_1),\ldots,(\obsfeaturenum,\featureind_\obsfeaturenum)$, which equal to 1. Thus, $\bm S \mathcal{P}_{\Omega'}(\mathring{\bm U}_1 \featurecoordinatevec)=\bm S \mathring{\bm U}_1 \featurecoordinatevec$.
\end{myitemize}%
\myitem\cmt{too many misses at eval \ra what to do}On the other hand, if
  $\obsfeaturenum:=|\Omega'| < \desiredrank$, it is not possible to
  identify $\featurecoordinatevec$ from $\testmissfeaturevec$. The
  extreme case would be when $\obsfeaturenum=0$. A
  natural estimate at such point can be the spatial average of the signal
  power $(1/N)\sum_n \tilde p_n$.
\end{myitemize}
\end{myitemize}
\section{Numerical tests}
\label{sec:numericalTest}
\begin{figure*}
 \centering\includegraphics[width=1.0\textwidth]{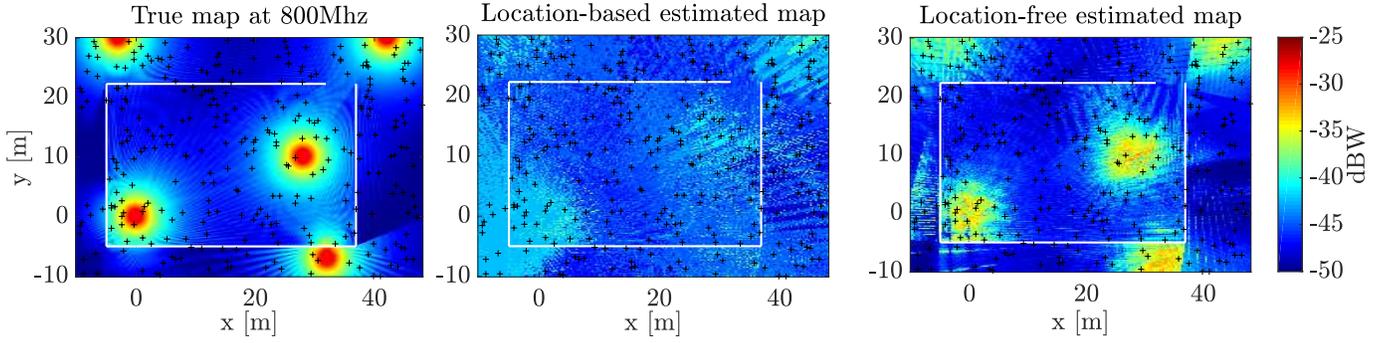}  
\caption{(left) True map, (middle) LocB  ($\regpar'=3.3\times 10^{-3}$, $\sigma'=0.5$ m), and (right) LocF ($\regpar=1.9\times 10^{-4}$, $\sigma=37$ m)
   estimated maps; $N=300$, $\sourcenum=5$, $B=20$
   MHz, and $K=10$. The black crosses indicate the sensor locations and the solid white lines represent the walls of the building.}
\label{f:2dtueAndEstMaps}
\end{figure*}
\begin{figure*}[t] 
\includegraphics[width=1.0\textwidth]{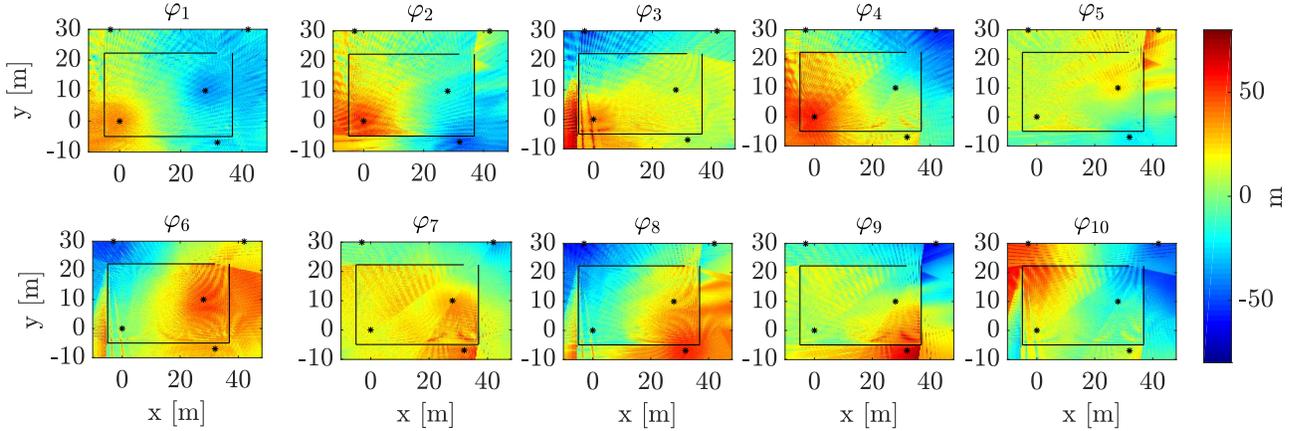} 
\caption{Maps of the $\featurenum=10$ LocF features with $\sourcenum=5$, $B=20$ MHz, and $K=10$. The solid black lines represent the walls of the
    building and the black stars represent the transmitter
    locations.} 
\label{f:smoothFvsnon-smoothB}
\end{figure*}
\begin{figure}[t]
\begin{center}
\includegraphics[width=1.0\columnwidth]{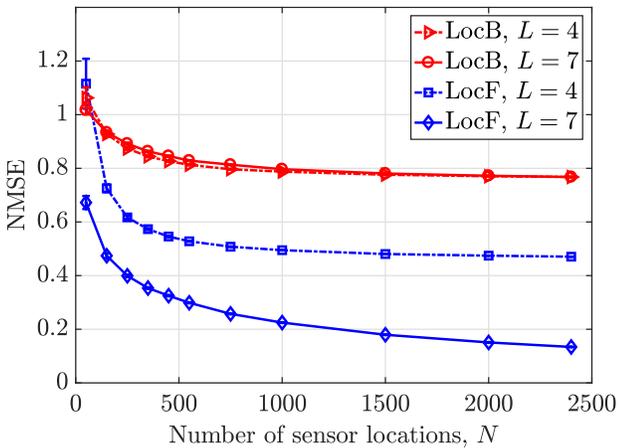}
\caption{Performance comparison
  between the LocF cartography ($\regpar=1.9\times 10^{-4}$, $\sigma=37$ m) and the LocB
  cartography ($\regpar'=3.3\times 10^{-3}$, $\sigma'=0.5$ m) with $B=20$
  MHz and $K=10$.} 
\label{f:FvsB}
\end{center} 
\end{figure}%
\begin{myitemize}
\myitem\cmt{Overview}This section evaluates the performance of
LocF cartography in presence of multipath, where 
localization algorithms cannot achieve accurate location estimates. 
\myitem\cmt{Simulation scenario}%
\begin{myitemize}%
\myitem\cmt{spatial region \ra building}To this end, the simulations are carried out in a 42 $\times$ 
27 m structure comprising several parallel vertical planes modeling the external and internal walls of a building, the latter is located in a 60 $\times$ 
40 m rectangular area $\mathcal{X}$.
\myitem\cmt{transmitters}%
\begin{myitemize}%
\myitem\cmt{number of transmitters}This area contains $\sourcenum$ active transmitters. Some of these are positioned inside the building, others outside. 
\myitem\cmt{Pilot signals}Matrix
$\pilotsigmat_n\in\cfield^{ \sourcenum \times K}$ containing the noisy
received pilot signals is generated according to \eqref{eq:rxsamples},
where $K$ is adjusted depending on $B$ to capture all the multipath
components.
\begin{myitemize}%
\myitem\cmt{pilot signal}For simplicity, the pilot signals are given
by\footnote{Amplitude units are such that a signal $x[k]=1,~\forall k,$
has power 1 W.} 
$\pilotsig_\sourceind[k]=\delta[k]$ which implies
that the rows of $\pilotsigmat_n\in\cfield^{ \sourcenum \times K}$
contain the impulse responses of the bandlimited channels between the
$\sourcenum$ transmitters and the $n$-th measurement location.
\myitem\cmt{channel}The channel $h_{\sourceind,n}[k]$ is generated
following \eqref{eq:discretetimechannel} with a carrier frequency of
800 MHz and  pilot channel bandwidth $B=1/T$. 
\myitem\cmt{noise term}The noise samples $\noisesig_{\sourceind,n}[k]$
are  independent normal
random variables with zero-mean and variance -70 dBm. 
\end{myitemize}
\begin{table*}
\normalsize
\renewcommand{\arraystretch}{1.25} 
\caption{Parameters used for the experiment in Fig.~\ref{f:LF_LBvsWallNumber}.}
\label{table:locFree_Based_par}
\centering \begin{tabular}{|c|c|c|c|c|c|}
\hline 
\multirow{2}{8em}&   $B$ (MHz)& $50$ & $100$ & $200$ & $700$ \\ \cline{2-6}
&$K$ & $25$ & $50$ & $100$ & $350$\\
\hline
\multirow{2}{8em}{ LocB}& $\sigma'$ (m)& $10.1$ & $8.9$ & $9$ & $7$ \\ \cline{2-6}
&$\regpar'$ & $1.8\times 10^{-3}$ & $9.1\times 10^{-4}$ & $7.1\times 10^{-4}$ & $2.1\times 10^{-4}$\\
\hline
\multirow{2}{8em}{ LocF} &$\sigma$ (m)& $27$  & $41$ & $53$ &$28$  \\\cline{2-6}
&$\regpar$ & $3.81\times 10^{-4}$ & $6.1\times 10^{-5}$ & $1.1\times 10^{-5}$ & $5\times 10^{-4}$\\ 
\hline
\end{tabular}
\end{table*}
\begin{figure*}[t] 
\includegraphics[width=1.0\textwidth]{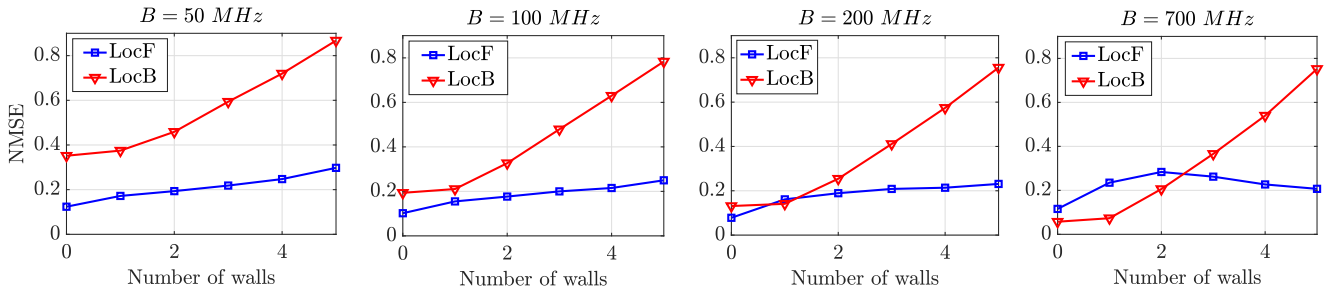} 
\caption{LocF and LocB map NMSE as a function of the number of walls
for different values of the bandwidth, $\sourcenum=5$,
$N=300$.} 
\label{f:LF_LBvsWallNumber}
\end{figure*}%
\end{myitemize}%
\myitem\cmt{channel}
\begin{myitemize}%
\myitem\cmt{Overview}Propagation adheres to the  \emph{Motley-Keenan
 multi-wall radio propagation model}~\cite{hosseinzadeh2017enhanced},
\myitem\cmt{Source of multipath}which accounts for the direct path,
up to 5 first-order wall reflections, and up to 5 wall-to-wall
second-order reflections. Remarkably, the model captures the impact
of the angle of incidence on the power of the reflected ray.
\myitem\cmt{C2M = loc ch}\arev{For simplicity, the C2M is chosen to be the channel where localization pilot
signals are transmitted. In practice, this is the case in the}
downlink of a cellular communication system such as LTE where the base stations
transmit both communication signals and localization pilots.
\end{myitemize}
\myitem\cmt{Measurements}

\begin{myitemize}
\myitem\cmt{sensor locations}To ensure that the measurements are
obtained in the far-field propagation region, sensor locations are
spread uniformly at random over $\bar{\mathcal{X}}$, which comprises
those points in $\mathcal{X}$ lying at least 3 wavelengths away from
all transmitters.
\myitem\cmt{sensor location number}Note that, although the number of
sensor locations is sometimes in the order of hundreds, this does not
mean that a large number of sensing devices must be used since each device may
gather measurements at tens or hundreds of spatial locations.
\myitem\cmt{Noise}The power measurement ${p}_n$ (measured in
dBW) of the C2M at position $\bm x_n$ is corrupted by additive noise
$\epsilon_n$ to yield $\tilde p_n = p_n + \epsilon_n$, where $\{\epsilon_n\}_{n=1}^N$ are independent
normal random variables with zero-mean and variance
$\sigma_\epsilon^2$. This variance is such that the signal-to-noise
ratio defined as $10\log_{10}(\bar p^2/\sigma_\epsilon^2)\approx 40$
dB, where $\bar{p}:=\int_{\bar{\mathcal{X}}}p(\bm x)d \bm
x/\int_{\bar{\mathcal{X}}}d \bm x$ is the spatial average of $p(\bm
x)$. This SNR is considered practical since the measurement noise power
$\sigma_\epsilon^2$ can be driven arbitrarily close to zero in
practice by
averaging over a sufficiently long time window.

\end{myitemize}
\end{myitemize}%

\myitem\cmt{Evaluation}Quantitative evaluation will
compare the normalized mean square error (NMSE) defined as: 
\begin{align}
\label{eq:nmsedef}
\text{NMSE}=\frac{\mathbb{E}\{ \vert p(\bm{x})-\hat{p}_{\pilotsigmat}(\bm{Y}(\bm x), \mathcal{T}) \vert ^2 \}}{\mathbb{E}\{ \vert
p(\bm{x})-\bar{p} \vert^2 \}},
\end{align} where $\hat{p}_{\pilotsigmat}(\bm{Y}(\bm x), \mathcal{T})$ (measured in dBW) denotes the
result of evaluating the map constructed from the training set
$\mathcal{T} := \lbrace
(\bm{Y}_{n},\tilde{p}_n)\rbrace _{n=1}^N$ at
the location $\bm x$, where $\pilotsigmat(\bm x)$ comprises the
received pilot signals at $\bm x$. The denominator
in \eqref{eq:nmsedef} normalizes the square error of the considered
algorithm by the error incurred by the best data-agnostic estimator,
which estimates the spatial average $\bar{p}$ at all points. Thus, the
adopted performance metric  is higher
than traditional NMSE, meaning that it is more challenging to obtain lower values. Furthermore $\mathbb{E}\lbrace \cdot
\rbrace $ denotes the expectation over the sensor locations and~noise.  
\end{myitemize}
\begin{figure}[t]
\begin{center}
\includegraphics[width=1.0\columnwidth]{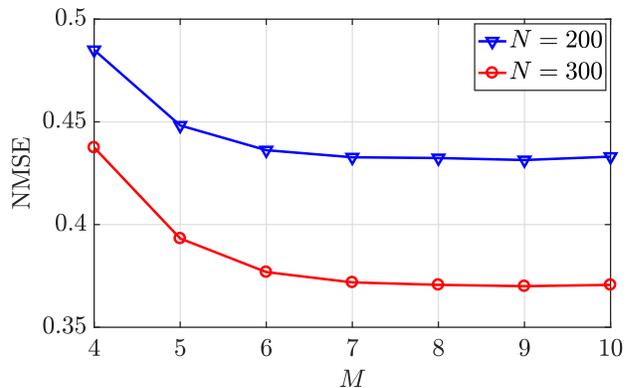}
\caption{ LocF estimated map NMSE for different values of number of features $\featurenum$ and sensor locations $N$, with $\sourcenum=5$, $B=20$ MHz, $K=10$, $\regpar=1.9\times 10^{-4}$, and $\sigma=37$ m.} 
\label{f:fdiffM}
\end{center} 
\end{figure}
\begin{figure*}[h!]
\centering
\begin{subfigure}{\textwidth}
\centering\includegraphics[width=0.85\columnwidth]{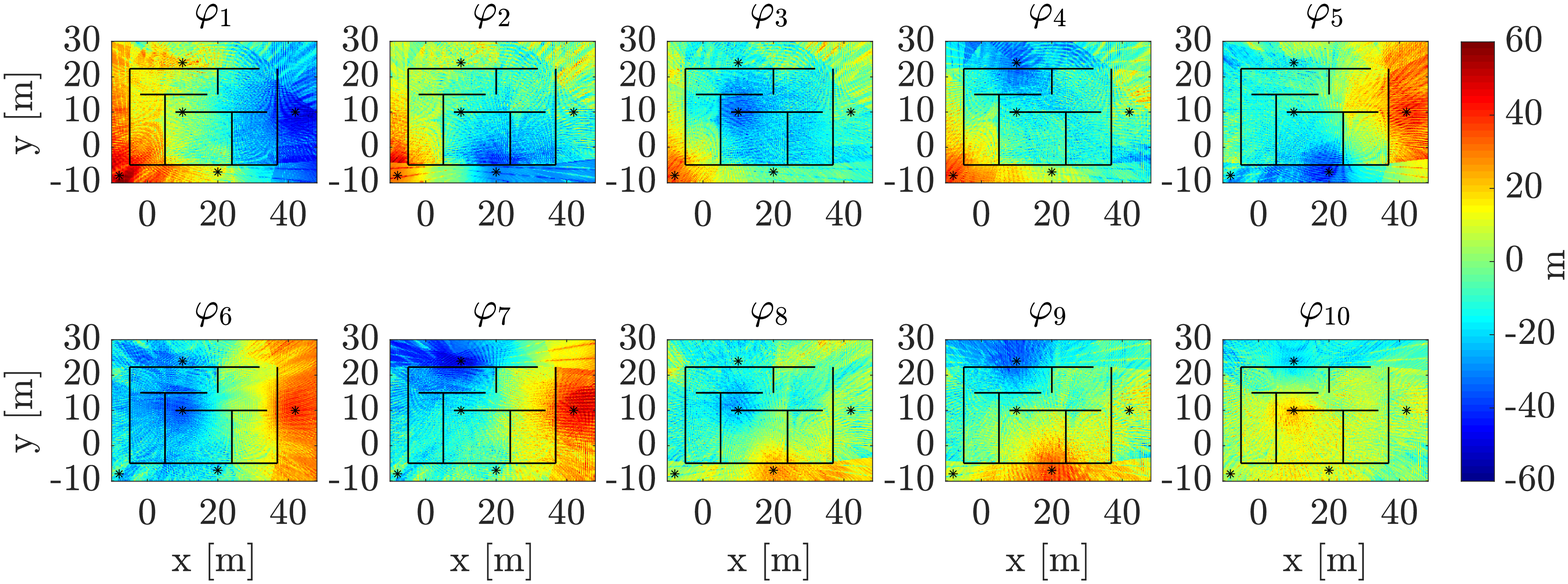}
\caption{}
\end{subfigure} \\
\begin{subfigure}{\textwidth}
\centering\includegraphics[width=0.85\columnwidth]{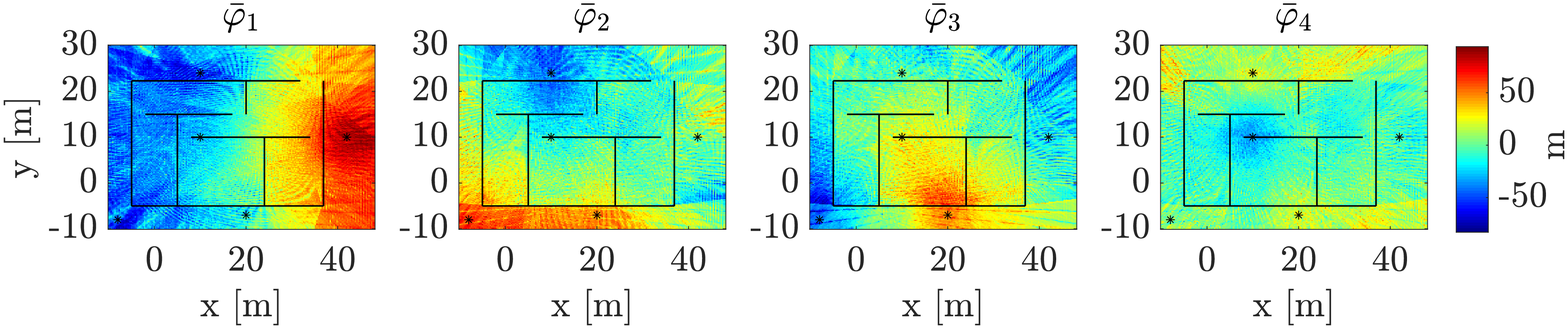}
\caption{}
\end{subfigure}
\caption{Maps of: (a) all the $\featurenum=10$ features and (b)
$\newfeaturenum=4$ reduced features with $\sourcenum=5$, $B=20$ MHz, and $K=10$. The solid black
lines represent the walls of the building and the black stars
represent the transmitter locations. The maps are obtained by
representing the value of the feature at every location in
$\mathcal{X}$.}
\label{f:pca}
\end{figure*}
\subsection{LocF vs. LocB}
\label{sec:4loc_free_and_based}
 \cmt{overview}
\cmt{LocF}
\begin{myitemize}
\myitem\cmt{features}
\begin{myitemize}
\myitem\cmt{Overview}To avoid the need for  synchronization between
transmitters and sensors, the LocF algorithm utilizes the features
in \eqref{eq:centerofmass}, which additionally provide robustness to
multipath and evolve smoothly over space; see
Sec.~\ref{sec:featureExtr}.
\myitem\cmt{Center of mass \ra distances}Since this center of mass can
be thought of as a lag, it is scaled by the sampling period $T$
and speed of light $c$ to obtain the corresponding range difference,
i.e.:
\begin{align}
\label{eq:comfeatures}
\begin{split}
\featurevec_n:=Tc~&[ 
\com\sensorsourcesourcenot{n}{1}{2},
\com\sensorsourcesourcenot{n}{1}{3},
\ldots,
\com\sensorsourcesourcenot{n}{1}{\sourcenum}, \\ &  
\com\sensorsourcesourcenot{n}{2}{3},
\ldots,
\com\sensorsourcesourcenot{n}{\sourcenum-1}{\sourcenum}
]^\top.
\end{split}
\end{align}
\end{myitemize}
\myitem\cmt{kernel learning}
\begin{myitemize}
\myitem\cmt{KRR}Using these features, the LocF algorithm uses the
kernel ridge regression technique in Sec.~\ref{sec:kernel} 
\myitem\cmt{kernel}with Gaussian radial basis functions with parameter
$\sigma$. The reason is that this \emph{universal kernel} is capable
of approximating arbitrary continuous functions that vanish at
infinity~\cite{micchelli2006universal}.
\end{myitemize}
\end{myitemize}
\cmt{LocB}On the other hand,
\begin{myitemize}
\myitem\cmt{features\ra localization}for LocB cartography, the feature
 vector $ \featurevec_n=\hbm x_n\in\rfield^2 $ comprises estimates of the
spatial coordinates of the $n$-th sensor location obtained by the 
\begin{myitemize}%
\myitem\cmt{loc.algorithm}\emph{iterative re-weighting squared range difference-least squares}
(IRWSRD-LS) algorithm~\cite{ismailova2015improved}, which features
 state-of-the-art localization performance.  
\myitem\cmt{features for loc. algo}This algorithm is applied over TDoA
 features extracted from $\{{ \pilotsigmat}_n\}_{n=1}^N$
through \eqref{eq:tdoafeat}.  At the $n$-th sensor location, these
features
$\{\hat{\tdoa}\sensorsourcesourcenot{n}{1}{\sourceind'}\}_{\sourceind'=2}^{\sourcenum}$
comprise the TDoA between a reference base station and the remaining
$\sourcenum-1$ base stations. 
\myitem\cmt{Not between all pairs of base stations, why}Enlarging this set by including 
TDoA measurements
$\hat{\tdoa}\sensorsourcesourcenot{n}{\sourceind}{\sourceind'}$ with
$\sourceind\neq 1$ would not be beneficial for the estimation
performance as discussed in~\cite{kaune2012accuracy}. The reason is
the redundancy inherent to TDoA features described in Sec.~\ref{sec:reducedFeat}.
\end{myitemize}
\myitem\cmt{kernel learning}To ensure a fair comparison, LocB utilizes
the same function learning algorithm as LocF; see Sec.~\ref{sec:kernel}.
\begin{myitemize}
\myitem\cmt{KRR}Specifically,
given $\left\lbrace
(\hbm{x}_{n},\tilde{p}_n)\right\rbrace _{n=1}^N$,
the map  is estimated as $\hat
p(\hbm{x})=\bm{\kappa}'^{\top}(\hbm{x})\hbm{\beta}$ where
\begin{myitemize}%
\myitem\ $\bm{\kappa}'(\hbm{x}):=\left[\kappa'(\hbm{x},\hbm{x}_{1}),\ldots,\kappa'(\hbm{x},\hbm{x}_{N})
  \right]^{\top}$, \myitem\ $\hbm{\beta}:= (\bm{K}'+\regpar' N
\textbf{I}_{N})^{-1} \tilde{\bm{p}}$, \myitem\ and $\bm{K}'$ is an $N
\times N$ matrix with $(n,n')$-th entry
$\kappa'(\hbm{x}_{n},\hbm{x}_{n'})$ and $\kappa'$ is a Gaussian radial
basis function with parameter $\sigma'$.
\end{myitemize}%
\myitem\cmt{special case}In this way, this benchmark LocB algorithm coincides
with those in
\cite{romero2017spectrummaps,bazerque2013basispursuit}
when a power map must be estimated on a single frequency and with a
single kernel.
\end{myitemize}
\end{myitemize}%
\cmt{reg params}In all experiments, the values of $\regpar$, $\regpar'$,
$\sigma$, and $\sigma'$ used by the LocF and LocB schemes were tuned to
approximately yield the lowest NMSE.
\cmt{Experiment 1: estimated maps and visual comparison with true
map\ra   Fig. \ref{f:2dtueAndEstMaps} and Fig.~\ref{f:smoothFvsnon-smoothB}.}

\begin{myitemize}
\myitem\cmt{ground truth}Fig. \ref{f:2dtueAndEstMaps} (left) depicts the true map generated through the
  multi-wall model, where
    \begin{myitemize}
  \myitem\cmt{sensor locations}the black crosses indicate the sensor
  locations
  \myitem\cmt{walls}and the solid white lines represent the walls of the building.
 \end{myitemize}%
  \myitem\cmt{estimated maps}The middle
  and right panels respectively show the LocB and
  LocF map estimates,  obtained by placing a
  query sensor at every location.
  \myitem\cmt{Observation}It is observed that
    the quality of the LocF estimate is considerably higher than that of the
  LocB estimate. 
  \myitem\cmt{cause \ra due to multipath}The cause for the poor
    perfomance of the LocB algorithm is that the location estimates evolve in
    a non-smooth fashion across space, and attempting to learn the C2M from such non-smooth features is more challenging; see  Figs.~\ref{f:locEst}c
    and~\ref{f:locEst}d and the discussion in
    Sec.~\ref{sec:introduction}. To illustrate how the LocF approach
    alleviates this issue, Fig.~\ref{f:smoothFvsnon-smoothB} depicts
    the features used by the LocF estimator across
    $\mathcal{X}$. Specifically, if $\featurevec(\bm x)$ denotes the
    feature vector, obtained as in  \eqref{eq:comfeatures} for
    location $\bm x$, then the $\featureind$-th panel titled $\varphi_\featureind$ in
    Fig.~\ref{f:smoothFvsnon-smoothB} corresponds to the
    $\featureind$-th entry of  $\featurevec(\bm x)$ for each $\bm
    x\in \mathcal{X}$. It
    is observed~that the evolution of these proposed features across
    space is significantly smoother 
    than the one in Figs.~\ref{f:locEst}c and~\ref{f:locEst}d. 
\end{myitemize}%
\cmt{Experiment 2: NMSE curves LocF vs LocB\ra Fig.~\ref{f:FvsB}}
\begin{myitemize}%
\myitem \cmt{figure}A quantitative comparison is provided in Fig.~\ref{f:FvsB}, which shows the
NMSE as a function of the number of sensor locations $N$ for
 $\sourcenum=4$ and $7$ transmitters. The
error bars delimit intervals of 6 standard deviations of the NMSE across
the  200 independent Monte Carlo runs.
\myitem\cmt{Observation}It is observed that, with high significance, the
 proposed LocF cartography scheme outperforms its LocB counterpart for
 both values of $\sourcenum$ provided that the number of measurement
 locations is roughly larger than~150.
\end{myitemize}

\cmt{Experiment 3: NMSE curves LocF vs LocB as function of the number of walls}
\begin{myitemize}
\myitem\cmt{fundemental limit}The rest of the section studies the
 impact of multipath on the LocF and LocB cartography approaches by
 varying the number of
 walls.  \myitem\cmt{figure}Fig.~\ref{f:LF_LBvsWallNumber} shows the
 NMSE as a function of the number of walls for different values of
 $B$. The parameters used for both LocF and LocB schemes are listed in
 Table~\ref{table:locFree_Based_par}. The NMSE is obtained by also
 averaging over wall locations, which are confined to be in the
 positions of the walls in Fig.~\ref{f:smoothFvsnon-smoothB} plus an
 additional wall that divides the room in two. 

\myitem\cmt{observation}
\begin{myitemize}
\myitem{}As expected, for all the simulated values of $B$, the
 performance of both LocF and LocB schemes is degraded (yet more
 severely in LocB) as the number of walls increases. Moreover, the
 performance of the LocB improves significantly with the bandwidth,
 since a higher bandwidth allows a more accurate estimation of the
 TDoA. This is because multipath components arriving within a time
 interval of length $T =1/B$ cannot be resolved; see
 Sec.~\ref{sec:featureExtr} and references therein.  %
\myitem{}As intuition predicts, when multipath is sufficiently low and the
 bandwidth is sufficiently high, LocB cartography outperforms LocF. 
\myitem{}It is remarkable that LocF cartography exhibits robustness to
multipath since the NMSE remains approximately constant even for a
significant increase of multipath.
\end{myitemize}
\end{myitemize}
\subsection{Feature Design}
\label{sec:4pca}
\begin{figure}[t]
\begin{center}
\includegraphics[width=1.0\columnwidth]{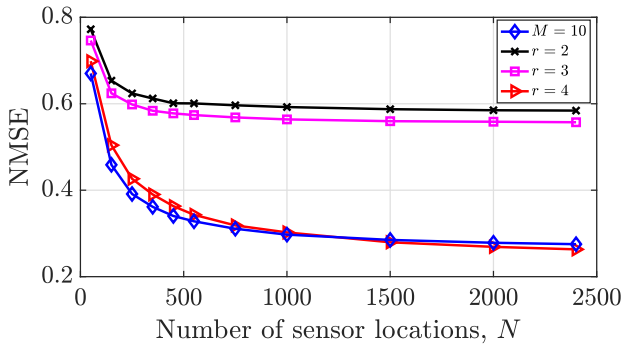}  
\caption{Estimated map NMSE with reduced features for different $\newfeaturenum$
and without reduced features; $\sourcenum=5$, $B=20$ MHz, $K=10$, $\regpar=1.6\times 10^{-3}$, and
$\sigma=25$ m.}
\label{f:pcaonMap}
\end{center}%
\end{figure}%
\cmt{overview}This section provides empirical support for the findings
in Sec.~\ref{sec:reducedFeat}. From now on, all experiments will
involve only the LocF estimator.
\cmt{Experiment 1: effect of feature number\ra Fig.~\ref{f:fdiffM}}%
\begin{myitemize}%
\myitem\cmt{for loc-free}The first experiment investigates the impact
of the number of features, which in all previous simulations was equal
to $\featurenum=\sourcenum(\sourcenum-1)/2$.
\begin{myitemize}%
\myitem\cmt{Figure}To this end, Fig.~\ref{f:fdiffM} shows the NMSE as
a function of the number $\featurenum$ of features for two different
numbers $N$ of sensor locations. The expectation operators
in \eqref{eq:nmsedef} also average  with respect to all choices of
$\featurenum$ features out of the ${\sourcenum(\sourcenum-1)/2}$. 
\begin{myitemize}
\myitem\cmt{observation}As observed, the NMSE improves  from $\featurenum=4$ to
roughly $\featurenum=7$ features,  and remains approximately the same for 
$\featurenum\geq 7$.
\myitem\cmt{explanation}Although this effect may look counter-intuitive at first
glance, this is a common phenomenon in machine learning related to the
bias-variance trade-off~\cite{cherkassky2007} and the curse of
dimensionality~\cite{bishop2006,cherkassky2007}; see
Sec.~\ref{sec:composition}. Clearly, this effect motivates 
the feature  dimensionality reduction techniques proposed in Sec.~\ref{sec:reducedFeat}.
\end{myitemize}
\end{myitemize} 
\end{myitemize}
\cmt{Experiment 2: dimensionality reduction\ra Figs.~\ref{f:pca} and Fig.~\ref{f:pcaonMap}}The rest of this section corroborates the merits
of such techniques. A more challenging scenario with more walls will
be considered. 
\begin{myitemize}%
\myitem\cmt{rotated features capture most of variance}The first step
is to determine the number of reduced features to be
used. It can be seen that  $\newfeaturenum=4$ in~\eqref{eq:retainedvar}
retains at least $\retainedvar = 99 \%$ of the variance of the
features in all tested scenarios. Thus, in principle, a map can be
learned using the reduced features $\featurecoordinatevec_n:=\bm
U_1^\top \featurevec_n\in \rfield^4$ without meaningfully sacrificing
estimation performance. 
\myitem\cmt{reduced  feats preserve smoothness}Before corroborating
that this is actually the case, it is instructive to visualize the
aforementioned reduced features across space.  Fig.~\ref{f:pca}a
portrays the maps of the $\featurenum=10$ original features, which
correspond to the entries of $\featurevec(\bm x)$; see
Sec.~\ref{sec:4loc_free_and_based}. On the other hand, the  panels of
Fig.~\ref{f:pca}b depict the reduced features over space, i.e., the 4
entries of the vector $\featurecoordinatevec(\bm x):=\bm
U_1^\top \featurevec(\bm x)$ for each $\bm x \in \mathcal{X}$. These
figures reveal that the reduced features inherit the spatial
smoothness of the original features. 

\myitem\cmt{the 4 proposed features attain same performance as all
features}To quantify the impact of reducing the dimensionality of the
feature vectors, Fig.~\ref{f:pcaonMap} compares the NMSE of the LocF
map estimate that relies on the original features ($\featurenum=10$)
with the one that relies on the reduced features
($\desiredrank=2,3,4$). As observed, using just the 4 reduced features
attains a similar performance to the estimator built on the 10
original features. This is expected given  the
bias-variance trade-off mentioned earlier. 
\myitem\cmt{Rel. to Fig.~\ref{f:fdiffM}}At this point, it might seem
that the effects observed in Fig.~\ref{f:fdiffM} contradict those of 
Fig.~\ref{f:pcaonMap} since in the former the NMSE is lower when 10
features are used relative to the case where only 4 are used. However,
that should not be concluded since the features in
Fig.~\ref{f:fdiffM} correspond to the entries of $\featurevec_n$
(see \eqref{eq:comfeatures}) whereas the features in
Fig.~\ref{f:pcaonMap}  correspond to the entries of
$\featurecoordinatevec_n:=\bm U_1^\top \featurevec_n$. 
\end{myitemize}%
\subsection{LocF cartography with Missing Features}
\label{sec:4misses}
\begin{figure}[t]
\begin{center}%
\includegraphics[width=0.80\columnwidth]{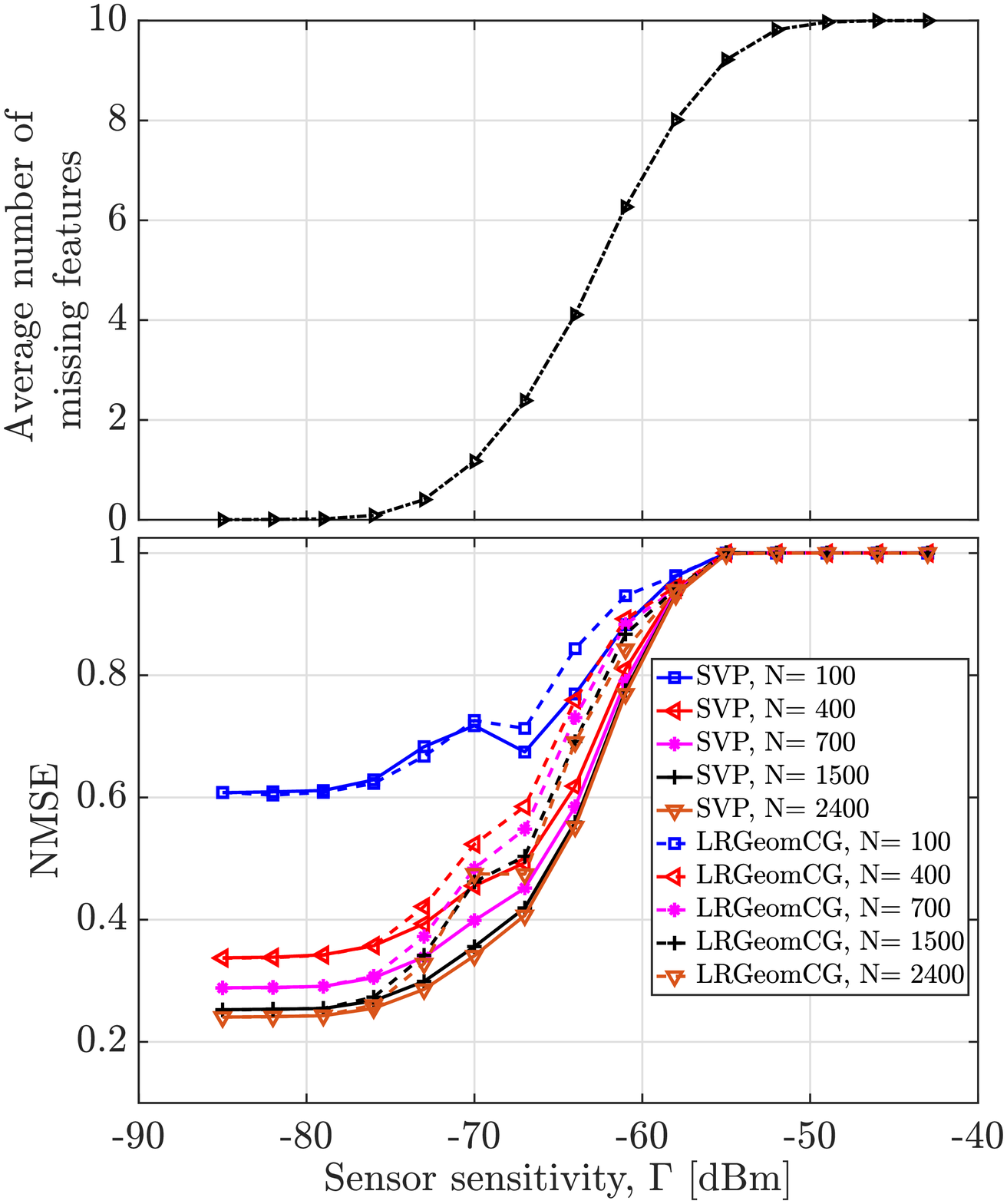}  
\caption{(top) Average number of missing features and (bottom) estimated map NMSE, both as a function of $\sensitivity$ with $\sourcenum=5$, $B=20$ MHz, $K=10$, $\regpar=1.9\times 10^{-4}$, $\evalregpar=5.42$,  and $\sigma=37$ m.}
\label{f:missesOnMap}
\end{center}%
\end{figure}%
\cmt{Experiment 1: Missing features \ra Fig.~\ref{f:missesOnMap}}This
section assesses the performance of the techniques developed in
Sec.~\ref{sec:missing} to cope with  missing features.

\begin{myitemize}
\myitem\cmt{generation of misses}A feature will be deemed missing at a given
sensor location if the received power of at least one of the two associated
pilot signals is below a sensitivity threshold  $\sensitivity$. 
\myitem\cmt{misses on map}The top panel of Fig.~\ref{f:missesOnMap} depicts
the average number of missing features as a function of
$\sensitivity$. The average is taken with respect to the sensor
locations and noise.
\myitem\cmt{NMSE}The bottom panel of Fig.~\ref{f:missesOnMap} shows the LocF map NMSE also as a function of $\sensitivity$. 
\begin{myitemize}%
\myitem\cmt{Implementation}The matrix completion problem
in \eqref{eq:manifoldopt} is solved with both SVP and LRGeomCG; the
implementation for the latter  is the one provided in the ManOpt
toolbox~\cite{boumal2014manopt}. 
\myitem\cmt{observation \ra both schemes}For higher values of
$N$, the performance of both algorithms is clearly strongly determined
by  the average number of missing features. 
\myitem\cmt{Degrad \ra much for LRGeomCG}SVP seems to outperform
LRGeomCG in terms of NMSE. Besides, the computation time of SVP is
roughly half the one of LRGeomCG. 
\end{myitemize}
\end{myitemize}
\section{Conclusions}
\label{sec:conclusions}
\begin{myitemize}
\myitem\cmt{Summary}Location-free (LocF) cartography has been proposed as an
alternative to classical location-based (LocB) schemes, which suffer a strong performance
degradation when multipath impairs the propagation of localization
pilot signals. The central idea is to learn a map as a function of
certain features of the localization pilot signals. Building upon this
approach, kernel-ridge regression was applied to estimate power maps
from these features. Practical issues addressed in the paper include
feature design, dimensionality reduction, and dealing with missing
features.  Simulations corroborate the merits of LocF cartography
relative to LocB alternatives.
\myitem\cmt{future work}Future research will include mapping other
channel metrics such as power spectral density (PSD) and channel gain,
as well as  developing distributed and online extensions. 
\end{myitemize}

\if\editmode1 
\onecolumn
\printbibliography
\else
\bibliography{\bibfilenames}
\fi
\end{document}